# Revealing the transient ionization dynamics and mode-coupling mechanisms of helicon discharge through a self-consistent multiphysics model

Jing-Jing Ma[1], Lei Chang[1*], Ming-Yang Wu[2], Hua Zhou[3,4], Yi-Wei Zhang[3,4], Ilya Zadiriev[5], Elena Kralkina[5], Shogo Isayama[6], Shin-Jae You[7]

[1]Fundamental Plasma Physics and Innovative Applications Laboratory, School of Electrical Engineering, Chongqing University, Chongqing 400044, China

[2]Institute of Plasma Physics, HFIPS, Chinese Academy of Sciences, Hefei 230031, China

[3]Science Island Branch of Graduate, University of Science and Technology of China, Hefei 230026, China

[4]School of Physics, Peking University, Beijing 100871, China

[5]Physical Electronics Department, Faculty of Physics, Lomonosov Moscow State University, GSP-1, Leninskie Gory, Moscow, 119991, Russian Federation

[6]Department of Advanced Environmental Science and Engineering, Kyushu University, 6-1 Kasuga-Kohen, Kasuga, Fukuoka 816-8580, Japan

[7]Applied Physics lab for PLasma Engineering (APPLE), Department of Physics, Chungnam National University, Daejeon 34134, Republic of Korea

*Email: leichang@cqu.edu.cn

**Abstract**

Helicon plasma sources play a central role in applications ranging from material treatment to space propulsion and fusion, yet the physical processes governing their ignition, transient ionization, and mode evolution remain incompletely understood. Here we develop a self-consistent, fully coupled multiphysics framework that integrates Maxwell's equations, electron energy transport, drift–diffusion kinetics, and heavy-species chemistry to capture the complete spatiotemporal evolution of helicon discharges. The model reproduces experimental measurements across pressure, magnetic field, and frequency ranges, and reveals a previously unresolved transient ionization stage characterized by a rapid density rise within $\sim 10^{-4}$ s, accompanied by a two-peak electron temperature structure that governs the formation of the dense plasma core. By tracking the RF power flow and field topology, we characterize the transient redistribution of RF energy during ignition. A short-lived phase of localized energy deposition accompanies the onset of ionization, followed by an evolution toward helicon-like field characteristics together with rapid density growth and profile restructuring. Systematic parametric scans further reveal the sensitivity of this mode-coupling process to gas pressure, magnetic field strength, and driving frequency. These results provide a unified picture of the ignition and mode-transition physics in helicon plasmas and establish a predictive tool for the design and optimization of RF plasma sources across space propulsion, manufacturing, and fusion technologies.

# 1. Introduction

Helicon plasmas are high-density magnetized discharges sustained by radio-frequency (RF) helicon waves and are valued for their high ionization efficiency, stable operation under low magnetic field and pressure, and electrode-less configuration [1-4]. These advantages have enabled their widespread adoption in plasma-based material processing, including surface modification [5], thin-film deposition [6-7], and semiconductor etching [8]. Beyond industrial applications, helicon plasmas are increasingly important in fields such as space propulsion and magnetic confinement fusion [9-13]. They serve as the core plasma source in systems including the Variable Specific Impulse Magnetoplasma Rocket (VASIMR) [14], helicon plasma thrusters (HPTs) [15-17], and double-layer helicon thrusters [18], where a magnetic nozzle formed by a divergent field accelerates ions to generate thrust [19-24]. More recently, helicon waves have been explored for heating, current drive, and active density control in fusion experiments [25-30].

Despite these diverse applications, the ionization and ignition mechanisms of helicon plasmas remain incompletely understood. Most studies focus on the steady state, whereas the early transient stage—during which ionization, heating, and wave coupling evolve rapidly—has received comparatively little attention [31-35]. Experiments have shown that plasma parameters change abruptly during ignition: optical emission spectroscopy revealed fast evolution of electron and hydrogen atom densities within the first few hundred microseconds [36], and Langmuir probe measurements indicated that quasi-steady conditions emerge only after ~250 μs [37]. Time-resolved laser spectroscopy has shown that downstream density can stabilize within several hundred microseconds [38], while pulsed experiments have reported transient electron heating above 8 eV and high-energy electron groups exceeding 40 eV [39]. Laser-induced fluorescence measurements have further revealed evolving ion velocity distributions and delayed formation of double layers [40-42]. Thomson scattering diagnostics have captured rapid temperature and density evolution [43], and recent work has identified multiple regimes in the relationship between power and density associated with mode transitions in W-mode discharges [44]. In parallel, numerical simulations have advanced the understanding of helicon plasma structure and dynamics. Multidimensional fluid models have captured wave propagation, power deposition, transport, and ionization processes [45], identified the influence of neutral depletion [46], and demonstrated parameter sensitivities associated with pressure, magnetic field, frequency, and input power [47]. Three-dimensional simulations have further revealed the interplay between ion pumping, RF heating, and collisional cooling, consistent with optical emission measurements [48]. These studies emphasize the central roles of RF power deposition, neutral depletion, and coupled ionization–heating dynamics in shaping transient behavior and mode transitions. Collectively, these studies indicate that helicon plasmas undergo a strongly non-equilibrium transient phase during ignition, and that its underlying physics remains unresolved—motivating the development of the present multiphysics model.

Nonetheless, significant gaps remain. Experimentally, it is challenging to measure spatiotemporally resolved ionization rates, wave fields, and power deposition, especially within the dense core region. Numerically, many models rely on simplifying assumptions or treat plasma transport, ionization kinetics, and electromagnetic fields in a loosely coupled manner, limiting their ability to capture nonlinear, multiscale transient dynamics. The quantitative mechanisms governing the rapid density buildup and evolving RF energy coupling during ignition therefore remain unclear. It is generally recognized that energy transfer is mediated by two principal wave modes—a weakly damped helicon wave and a strongly damped electrostatic Trivelpiece–Gould (TG) wave [49-51]—and that the TG wave plays a dominant role in power deposition and initial electron heating [4]. However, existing models often assume steady-state conditions or reduced dimensionality, hindering efforts to resolve the spatiotemporal wave–plasma coupling that drives early-time ionization.

To address these challenges, we develop a two-dimensional axisymmetric multiphysics coupled model based on the COMSOL Multiphysics platform. The framework integrates drift–diffusion electron transport, heavy-particle kinetics, Maxwell's equations, and key chemical processes, enabling the plasma to be treated self-consistently across the core and edge regions. The simulations reveal that helicon plasma ignition depends sensitively on external operating conditions, with neutral gas pressure, external magnetic field, antenna driving frequency, and RF power each shaping density growth, temperature evolution, and spatial power deposition in distinct ways. The model reproduces characteristic transient features—including rapid density rise, temperature overshoot, and evolving wave-field structure—and clarifies the mechanisms through which wave–particle interactions govern the nonequilibrium ionization process. By capturing multiscale transport and spatiotemporal power deposition, this approach bridges the gap between experimental diagnostics and predictive modeling, providing quantitative insight into helicon plasma kinetics and guiding the optimized design of plasma sources for propulsion, manufacturing, and fusion applications.

## 2. Physical Model and Numerical Scheme

### 2.1 Computational domain

To investigate the spatiotemporal evolution of ionization in the helicon plasma, we refer to a highly repetitive and advanced device, i.e. Peking University Plasma Test (PPT) device [52] (Figure 1). The PPT device developed at Peking University consists of a cylindrical vacuum chamber with inner diameter of approximately 500 mm and total length of 1 m. A pair of external Helmholtz coils is installed to generate a uniform axial magnetic field ranging from 0 to 2000 G [53,54]. In the central region of the chamber, a quartz discharge tube with diameter of 150 mm and length of 400 mm is placed. A single-loop RF antenna operating at the industrial frequency of 13.56 MHz is used to couple electromagnetic power into the plasma. For spatially resolved diagnostics, a radially movable Langmuir probe is mounted at the central plane of the chamber to measure plasma density, floating potential, and fluctuation characteristics. In addition, a fast-framing imaging window is installed at the downstream end, enabling direct observation of the plasma evolution. Referring to this device, we develop a two-dimensional axisymmetric multiphysics model based on the COMSOL Multiphysics platform. Figure 2 shows the computational domain with the externally applied magnetic field included (background streams). The computational domain is divided into three subregions: the plasma region (D1), the antenna region (D2), and the surrounding vacuum (D3). In D1, the grounded discharge tube has diameter of 15 cm and length of 40 cm, whereas the overall chamber extends 100 cm axially with inner diameter of 25 cm. The working gas in D1 is treated as an ideal gas at 300 K, and D3 is assumed to be a perfect vacuum. The antenna region (D2) is modeled as a copper ring with radius of 5 cm and square cross section of 1 cm × 1 cm, located at z = 30 cm. To facilitate comparison with experimental diagnostics, a virtual probe is positioned at z = 75 cm to monitor the local plasma parameters.

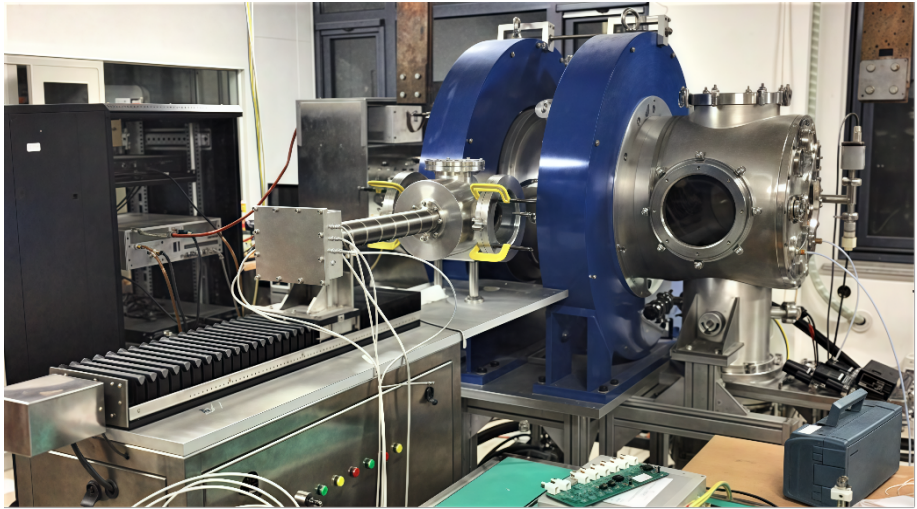

**Figure 1.** Photo of Peking University Plasma Test (PPT) device (reproduced with permission from Tian-Chao Xu, "Experimental study of turbulent transport on the PPT device and a novel magnetic field diagnostic method of Field-reversed Configuration (FRC)", Peking University, 2020, Ref. [52]).

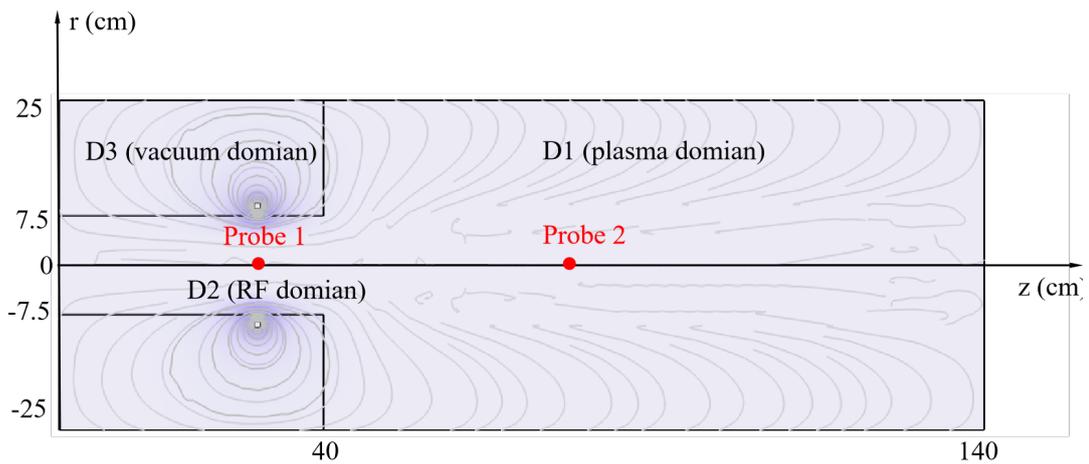

**Figure 2.** Computational domain referring to Peking University Plasma Test (PPT) device.

**2.2 Physical model**

To describe the plasma generation, power deposition, and transport characteristics in the helicon discharge, a multiphysics coupled model is established, incorporating the electron drift-diffusion equation, the heavy-particle transport equation, and the electromagnetic field equations, while neglecting electron convection [55] and gas temperature gradients for simplicity [56]. The electron transport is described using a drift-diffusion framework that captures the spatiotemporal evolution of both electron density and energy density. The formulation incorporates drift motion under electric fields, diffusion driven by density gradients, and collisional source terms, while accounting for energy losses and power absorption in the discharge. The governing equations can be expressed as follows [57]:

$$\begin{cases} \dfrac{\partial n_e}{\partial t} + \nabla \cdot \mathbf{\Gamma}_e = R_e \\ \dfrac{\partial}{\partial t}(n_\varepsilon) + \nabla \cdot \mathbf{\Gamma}_\varepsilon + \mathbf{E} \cdot \mathbf{\Gamma}_\varepsilon = S_{en} \end{cases} \quad (1)$$

Here, $n_e$ denotes the electron density, $n_\varepsilon$ represents the electron energy density, and $\mathbf{\Gamma}_e$ and $\mathbf{\Gamma}_\varepsilon$ correspond to the fluxes of electrons and energy, respectively. $S_{en}$ represents the energy loss due to inelastic collisions, $\mathbf{E}$ is the electric field, and $R_e$ denotes the electron source term, which accounts for the increase in electron density caused by ionization and excitation processes. It can be expressed as:

$$R_e = \sum_{j=1}^{M} v_j k_j n_e n_{t,j} \quad (2)$$

In the equation, $v_j$ is the electron velocity coefficient, $k_j$ represents the rate constant of the $j$-th reaction, $n_e$ is the electron density, and $n_{t,j}$ denotes the number density of the target species involved in the $j$-th reaction. For a neutral gas in the ground state, the number density is given by

$$\begin{cases} n_{t,j} = x_j N_n \\ N_n = \dfrac{P_A}{k_B T_g} \end{cases} \quad (3)$$

Here, $x_j$ represents the mole fraction. The energy loss term $S_{en}$ primarily arises from inelastic collision processes, and it can be expressed as:

$$S_{en} = \sum_j n_e n_n k_j \Delta \varepsilon_j \quad (4)$$

Here, $n_n$ denotes the neutral particle density, and $\Delta \varepsilon_j$ represents the energy loss corresponding to the $j$-th reaction. This term represents the loss of electron energy resulting from excitation, ionization, and other inelastic processes. In the drift–diffusion formulation, the electron particle flux is written as [58]:

$$\mathbf{\Gamma}_\mathbf{e} = -n_e(\mu_e \mathbf{E}) - D_e \nabla n_e \quad (5)$$

where $\mu_e$ is the electron mobility and $D_e$ is the electron diffusivity. The definition of the electron density flux also adopts the assumption from Ref. [59]: the electron temperature gradient term is neglected (i.e., the electron heat conduction process is not considered). This is justified because the electron temperature exhibits a relatively uniform spatial distribution within the plasma [48]. Due to the presence of background magnetic field, the plasma exhibits anisotropy. In the right-handed cylindrical coordinate system $(r,\varphi,z)$, the electron mobility can be expressed in a three-dimensional tensor form as follows:

$$\mathbf{\mu}_e^{-1} = \begin{bmatrix} \dfrac{1}{\mu_{dc}} & -\mathbf{B}_z & \mathbf{B}_\varphi \\ \mathbf{B}_z & \dfrac{1}{\mu_{dc}} & -\mathbf{B}_r \\ -\mathbf{B}_\varphi & \mathbf{B}_r & \dfrac{1}{\mu_{dc}} \end{bmatrix} \tag{6}$$

Here, $\mu_{dc}$ represents the direct-current electron mobility in the absence of a magnetic field. The electron energy distribution function (EEDF) in the plasma source region is approximated as a Maxwellian distribution. Experimental results from Ref. [60] also confirm that the deviation of EEDF from Maxwellian distribution in the source region is not significant. When the EEDF follows a Maxwellian distribution, integration of the Boltzmann equation yields the interrelations among the transport coefficients [61], from which the electron diffusivity, energy mobility, and energy diffusivity can be derived respectively as follows:

$$\begin{cases} \mathbf{D}_e = \mathbf{\mu}_e T_e \\ \mathbf{D}_\varepsilon = \mathbf{\mu}_\varepsilon T_e \\ \mathbf{\mu}_\varepsilon = \dfrac{5}{3}\mathbf{\mu}_e \end{cases} \tag{7}$$

Since the actual plasma in this model may contain many neutral particles and excited particles, the direct solution of the Maxwell–Stefan equations requires considering friction coefficients, pressure and temperature gradients, and drift transport. This makes the system particularly difficult to handle. Therefore, the present model adopts the Fick approximation to simplify the formulation, making it more computationally tractable and easier to obtain solutions. Assuming that there are $Q$ species of heavy particles, indexed by $k = 1, \ldots, Q$, and $N$ chemical reactions indexed by $j = 1, \ldots, N$, the mass fraction equation for non-electron species $\omega_k$ is written as follows [62]:

$$\rho \frac{\partial \omega_k}{\partial t} + \rho(\mathbf{u} \cdot \nabla)\omega_k = \nabla \cdot \mathbf{j}_k + R_k \tag{8}$$

where $\mathbf{j}_k$ is the diffusion flux, $\mathbf{u}$ is the bulk mass-averaged velocity, $\rho$ is the mixture density, and $\omega_k$ is the mass fraction of species $k$. The diffusion flux is expressed as:

$$\mathbf{j}_k = \rho \omega_k \mathbf{V}_k \tag{9}$$

where $\mathbf{V}_k$ is the diffusion velocity of species $k$. The definition of $\mathbf{V}_k$ depends on the selected diffusion model. If the diffusion model adopts the Fick-law approximation, a diffusion coefficient must be specified for each species. The diffusion velocity is defined as:

$$\mathbf{V}_k = D_{k,f} \frac{\nabla \omega_k}{\omega_k} + D_{k,f} \frac{\nabla M_n}{M_n} + D_k^T \frac{\nabla T}{T} - z_k \mu_{k,f} \mathbf{E} \tag{10}$$

where $D_{k,f}$ is the Fick diffusion coefficient. The ion mobility $\mu_{k,f}$ is expressed as:

$$\mu_{k,f} = \frac{qD_{k,f}}{k_B T} \tag{11}$$

The Fick diffusion coefficient $D_{k,f}$ is computed through the Einstein relation:

$$D_{k,f} = \frac{k_B}{q}\mu_{k,f} \tag{12}$$

The average molar mass $M_n$ is obtained from:

$$\frac{1}{M_n} = \sum_{k=1}^{Q} \frac{\omega_k}{M_k} \tag{13}$$

The chemical reaction source term for heavy species is given by:

$$R_k = M_k \sum_{j=1}^{N} \nu_{k,j} r_j \tag{14}$$

where $\nu_{k,j}$ is the stoichiometric coefficient, $r_j$ is the reaction rate of reaction $j$, and $M_k$ is the molar mass of species $k$. Since Eq. (8) consists of $Q-1$ species conservation equations together with the mass conservation constraint, the complete description of all heavy species satisfies:

$$\sum_{k=1}^{Q} \omega_k = 1 \tag{15}$$

In the helicon discharge, the total electric field can be decomposed into two parts. The first part is the time-varying electromagnetic field generated by the RF antenna, including the field induced by the associated eddy currents. The second part is the quasi-static electric field formed by ambipolar diffusion in the plasma. The time-harmonic electromagnetic component is governed by Ampère's law written in terms of the magnetic vector potential, whereas the quasi-static component is described by Poisson's equation in terms of the electrostatic potential. By rewriting Ampère's law in the frequency domain, the governing equation for the vector potential $\mathbf{A}$ can be expressed as

$$(j\omega\sigma - \omega^2 \varepsilon_0)\mathbf{A} + \nabla \times (\mu_0^{-1} \nabla \times \mathbf{A}) = \mathbf{J}_e \tag{16}$$

Here, $\mathbf{A}$ denotes the magnetic vector potential, $\sigma$ is the plasma conductivity, $\mu_0$ and $\varepsilon_0$ are the magnetic permeability and permittivity of free space, respectively, $\mathbf{J}_e$ represents the induced current density in the plasma. When solving the governing equations across different computational domains (antenna, vacuum, quartz tube, and the plasma region), different material properties are applied: constant material parameters are used outside the plasma region, whereas a tensor form is employed inside the plasma where magnetization effects are significant.

$$\boldsymbol{\sigma} = \begin{bmatrix} \dfrac{m_e(v_e + j\omega)}{n_e e^2} & -\dfrac{B_z}{n_e e} & \dfrac{B_y}{n_e e} \\ \dfrac{B_z}{n_e e} & \dfrac{m_e(v_e + j\omega)}{n_e e^2} & -\dfrac{B_x}{n_e e} \\ -\dfrac{B_y}{n_e e} & \dfrac{B_x}{n_e e} & \dfrac{m_e(v_e + j\omega)}{n_e e^2} \end{bmatrix}^{-1} \quad (17)$$

where $v_e$ is the electron collision frequency, and $\omega$ is the angular frequency of the RF field. The corresponding ohmic power dissipation can be expressed as [48]:

$$Q_{rh} = \tfrac{1}{2}\text{Re}(\mathbf{J}^* \cdot \mathbf{E}_{RF}) = \tfrac{1}{2}\text{Re}\left[(\sigma \mathbf{E}_{RF} + j\omega \mathbf{E}_{RF})^* \cdot \mathbf{E}_{RF}\right] \quad (18)$$

where $\text{Re}$ denotes the real part, $\mathbf{J}$ is the current density in the plasma, and $\mathbf{E}_{RF} = -j\omega \mathbf{A}$ presents the oscillating electric field. Since $\text{Re}\left[(j\omega\varepsilon \mathbf{E}_{RF})^* \cdot \mathbf{E}_{RF}\right] = 0$, the displacement current corresponds to a purely reactive component and does not contribute to real power absorption. Hence, only the real part of $\sigma$ contributes to resistive heating, and Eq. (18) can be rewritten as:

$$\begin{cases} Q_{rh} = \tfrac{1}{2}\sigma'(\omega)\mathbf{E}_{RF}^2 \\ \sigma'(\omega) = \text{Re}\,\sigma(\omega) \end{cases} \quad (19)$$

For a collisional plasma, the Drude model is adopted for the conductivity, expressed as

$$\sigma(\omega) = \frac{n_e e^2}{m_e(v_e - j\omega)} = \frac{n_e e^2}{m_e}\frac{v_e + j\omega}{v_e^2 + \omega^2} \quad (20)$$

The real part of the conductivity can be written as

$$\sigma'(\omega) = \frac{n_e e^2 v_e}{m_e(v_e^2 + \omega^2)} \quad (21)$$

Substituting Eq. (21) into Eq. (19), the final expression for power dissipation is obtained as

$$Q_{rh} = \tfrac{1}{2}\frac{n_e e^2 v_e}{m_e(v_e^2 + \omega^2)}|\mathbf{E}_{RF}|^2 \quad (22)$$

We shall mainly solve Eq. (1), Eq. (8), Eq. (16), and Eq. (18) in a coupled manner, i.e. in COMSOL Multiphysics framework, to obtain the spatial and temporal evolution of plasma parameters, particularly the ionization rate, electron density, and power deposition in helicon discharge. This will be done with the combinations of following chemical reactions.

In this model, the key collision and reaction processes among ground-state argon atoms (Ar), metastable argon atoms (Ars), and argon ions ($Ar^+$) are taken into account, as adopted in previous literature [63]. As listed in Table 1, the reaction set includes seven volumetric and two surface processes. The volumetric reactions consist mainly of elastic scattering between electrons and argon atoms, excitation from the ground state, superelastic collisions, and both direct and stepwise ionization. Among these, ground-state excitation and superelastic collisions are major channels for electron energy transfer, while stepwise ionization becomes the dominant mechanism sustaining high plasma density at low pressures. Direct ionization from the ground state governs the overall increase in electron density. Furthermore, Penning ionization ($Ars+Ars \rightarrow e+Ar+Ar^+$) provides an additional electron source even at relatively low electron energies, which helps to enhance the total ionization efficiency. In addition to the volumetric reactions, two surface processes are included: quenching of metastable argon atoms ($Ars \rightarrow Ar$) and recombination between ions and electrons at the wall ($Ar^++e \rightarrow Ar$). The volumetric reactions mainly determine plasma production and power deposition inside the discharge, while the surface reactions are responsible for particle loss and boundary effects near the wall.

**Table 1.** Argon collisions and chemical reactions.

| No. | Reaction Process | Type | $\Delta\varepsilon$(eV) |
|---|---|---|---|
| 1 | $e + Ar \Rightarrow e + Ar$ | Elastic collision | 0 |
| 2 | $e + Ar \Rightarrow e + Ars$ | Excitation | 11.5 |
| 3 | $e + Ars \Rightarrow e + Ar$ | Superelastic collision | -11.5 |
| 4 | $e + Ar \Rightarrow 2e + Ar^+$ | Ionization | 15.8 |
| 5 | $e + Ars \Rightarrow 2e + Ar^+$ | Step ionization | 4.24 |
| 6 | $Ars + Ars \Rightarrow e + Ar + Ar^+$ | Penning ionization | - |
| 7 | $Ars + Ar \Rightarrow Ar + Ar$ | Metastable quenching | - |
| 8 | $Ars \Rightarrow Ar$ | Surface recombination | - |
| 9 | $Ar^+ \Rightarrow Ar$ | Surface recombination | - |

Considering the dominant exchange mechanisms of electrons at the plasma–wall interface, the electron particle flux and electron energy flux at the wall are described by the following expressions:

$$\mathbf{n} \cdot \Gamma_e = \frac{1-r}{1+r}\left(\frac{1}{2}v_{e,\text{th}}n_e\right) - \frac{2}{1+r}(1-a)\left[\sum_p \gamma_p (\Gamma_p \cdot \mathbf{n}) + \Gamma_t \cdot \mathbf{n}\right] \qquad (23)$$

$$\mathbf{n} \cdot \Gamma_\varepsilon = \frac{1-r}{1+r}\left(\frac{5}{6}v_{e,\text{th}}n_e\right) - \frac{2}{1+r}(1-a)\left[\sum_p \gamma_p \bar{\varepsilon}_p (\Gamma_p \cdot \mathbf{n}) + \bar{\varepsilon}_t (\Gamma_t \cdot \mathbf{n})\right] \qquad (24)$$

Here, $\mathbf{n}$ is the outward unit normal vector of the wall surface, $v_{e,\text{th}}$ denotes the electron thermal velocity, and $r$ is the electron reflection coefficient. $\Gamma_p$ represents the flux of positive ion species $p$ incident on the wall, and $\gamma_p$ is the corresponding secondary electron emission coefficient. $\bar{\varepsilon}_p$ is the mean energy of secondary electrons, while $\Gamma_t$ and $\bar{\varepsilon}_t$ denote the thermionic emission flux and its mean energy, respectively. In this work, $r = 0$ is used to neglect electron reflection, and thermionic emission is omitted by setting $\Gamma_t = 0$.

A grounded boundary is applied, satisfying the zero-potential condition:

$$V = 0 \tag{25}$$

and the initial conditions are set as:

$$n_e = 1 \times 10^{16} \, \text{m}^{-3}, \quad T_e = 2 \, \text{eV} \tag{26}$$

where $n_e$ is the initial electron density, $T_e$ is the initial electron temperature. Numerical simulations are carried out in COMSOL Multiphysics platform. In our study, time-dependent solver based on the implicit backward differentiation formula with adaptive time stepping is adopted. The initial step size is set to $10^{-13}$ s, and the solver subsequently adjusts the time steps to capture the rapid transient evolution of the helicon discharge. The relative tolerance is maintained at $10^{-3}$ to ensure numerical stability throughout the calculation. Meanwhile, a locally refined mesh is applied in the discharge tube and antenna region, providing adequate spatial resolution for resolving both the electromagnetic field and particle transport. Based on the settings given above, the geometric and physical parameters of the PPT device are listed in Table 2 to ensure consistency between the simulation and experimental conditions.

**Table 2.** Dimensions and Parameters of the PPT Device.

| No. | Parameter | Value |
|---|---|---|
| 1 | Discharge tube diameter $d_{D1}$ | 15 cm |
| 2 | Discharge tube length $L_{D1}$ | 40 cm |
| 3 | Vacuum chamber axial length $L_v$ | 100 cm |
| 4 | Vacuum chamber inner diameter $d_v$ | 25 cm |
| 5 | RF antenna radius $R_a$ | 5 cm |
| 6 | Antenna cross-section | 1×1 cm² |
| 7 | Antenna position $probe_1$ | 30 cm |
| 8 | Langmuir probe position $probe_2$ | 75 cm |
| 9 | Working gas | Argon gas |
| 10 | Initial gas temperature $T_g$ | 300 K |
| 11 | RF frequency $f$ | 13.56 MHz |
| 12 | RF power $P_{RF}$ | ~2250 W |
| 13 | Neutral pressure $p$ | ~10 Pa |
| 14 | Axial uniform magnetic field $B_0$ | ~2000 G |

The present model improves upon previous helicon plasma simulations through directly coupling the electron drift–diffusion equations, heavy-particle transport, and electromagnetic fields within a single framework. This enables the simulation to resolve the startup, ionization, and RF power deposition processes in the PPT discharge with high spatial and temporal fidelity. An adaptive implicit scheme for time integration and a locally refined mesh are used to maintain numerical stability and capture the sharp gradients near the antenna and plasma core. As a result, our model reproduces experimental behavior more realistically and provides deeper insight into the transient evolution and ionization dynamics of helicon plasmas.

**2.3 Model validation**

To verify the reliability of the developed model and numerical scheme, our computation results are compared with both the experimental measurements obtained from the PPT device and the computation results from the Peking University Helicon Discharge (PHD) code [53]. The benchmark is done under the conditions of a background magnetic field of 1180 G and a neutral gas pressure of 0.35 Pa. Figure 3 compares the variations of the peak electron density with input power, neutral gas pressure, and magnetic field strength. It can be seen that our simulations agree quite well with the experimental results in both magnitude and trend. The agreement is even better than that from the PHD code, suggesting that our model captures the essential physics more accurately. The comparison of the maximum electron density under different neutral pressures shows good consistency when the pressure is below 0.6 Pa. Regarding the influence of magnetic field, the steady-state electron density grows rapidly with field strength when (B < 500) G, indicating a strong dependence on magnetic confinement. As the field continues to increase, the growth gradually levels off and eventually saturates. In general, our model and simulations capture the overall trend observed in the PPT experiments, though small deviations remain in absolute values. These differences may stem from uncertainties in experimental measurements and model simplifications. Therefore, the developed two-dimensional axisymmetric multiphysics coupled model reproduces the temporal evolution and spatial distribution of electron density in good consistency with both PPT experiment and PHD code, proving its reliability and fidelity.

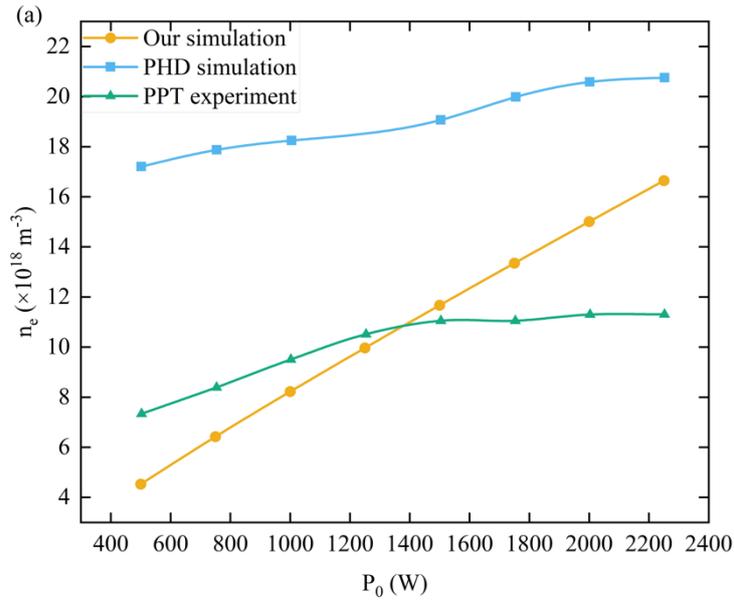

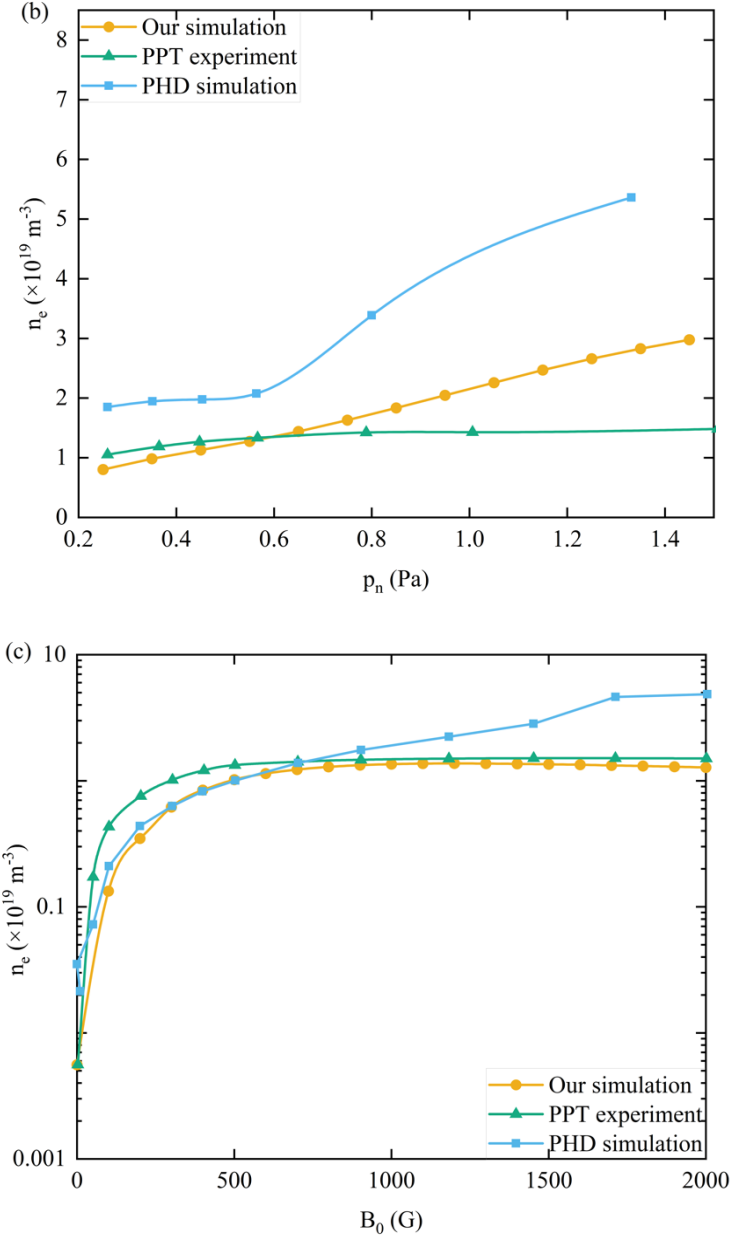

**Figure 3.** Benchmarked results of our model against experimental measurements on PPT and computation results of PHD code: relationships between the maximum electron density and (a) input power $P_0$, (b) neutral pressure $p_n$, and (c) magnetic field $B_0$.

## 3. Results and Discussion

### 3.1 Spatiotemporal evolution of helicon discharge

Helicon discharges exhibit intrinsically transient and spatially structured behavior during their early evolution, driven by the interplay between rapid ionization, cross-field diffusion, and electromagnetic power deposition. To resolve this non-equilibrium phase, we performed time-dependent simulations using an RF power of 1500 W, an axial magnetic field of 1180 G, and a neutral argon pressure of 0.5 Pa—conditions representative of the baseline PPT experiment [52,53]. The evolution of the electron density and electron temperature was examined both locally (at r = 0 cm, z = 75 cm, corresponding to Probe 2 in Fig. 2) and globally through spatiotemporal maps. Figures 4–7 summarize the corresponding dynamics.

- **Early ignition and radial migration of ionization front**

Figure 4 illustrates the temporal evolution of the on-axis electron density together with snapshots of its two-dimensional spatial structure at four representative times: I: $7.94 \times 10^{-8}$ s, II: $7.94 \times 10^{-5}$ s, III: $3.16 \times 10^{-4}$ s, IV: 0.01 s.

At the earliest moment (stage I), the RF field penetrates the weakly ionized gas with negligible shielding, and the first electrons are produced near the quartz wall close to the antenna. As ionization proceeds (stage II), the density increases rapidly and the ionization front expands inward. By $\sim 3 \times 10^{-4}$ s (stage III), the high-density region shifts toward the symmetry axis, accompanied by a gradual change in shielding and transport characteristics as the discharge develops. By 0.01 s, both radial and axial density profiles are close to steady state (stage IV). The density saturates as ionization balances transport and wall losses, forming the familiar on-axis density peak characteristic of helicon discharges.

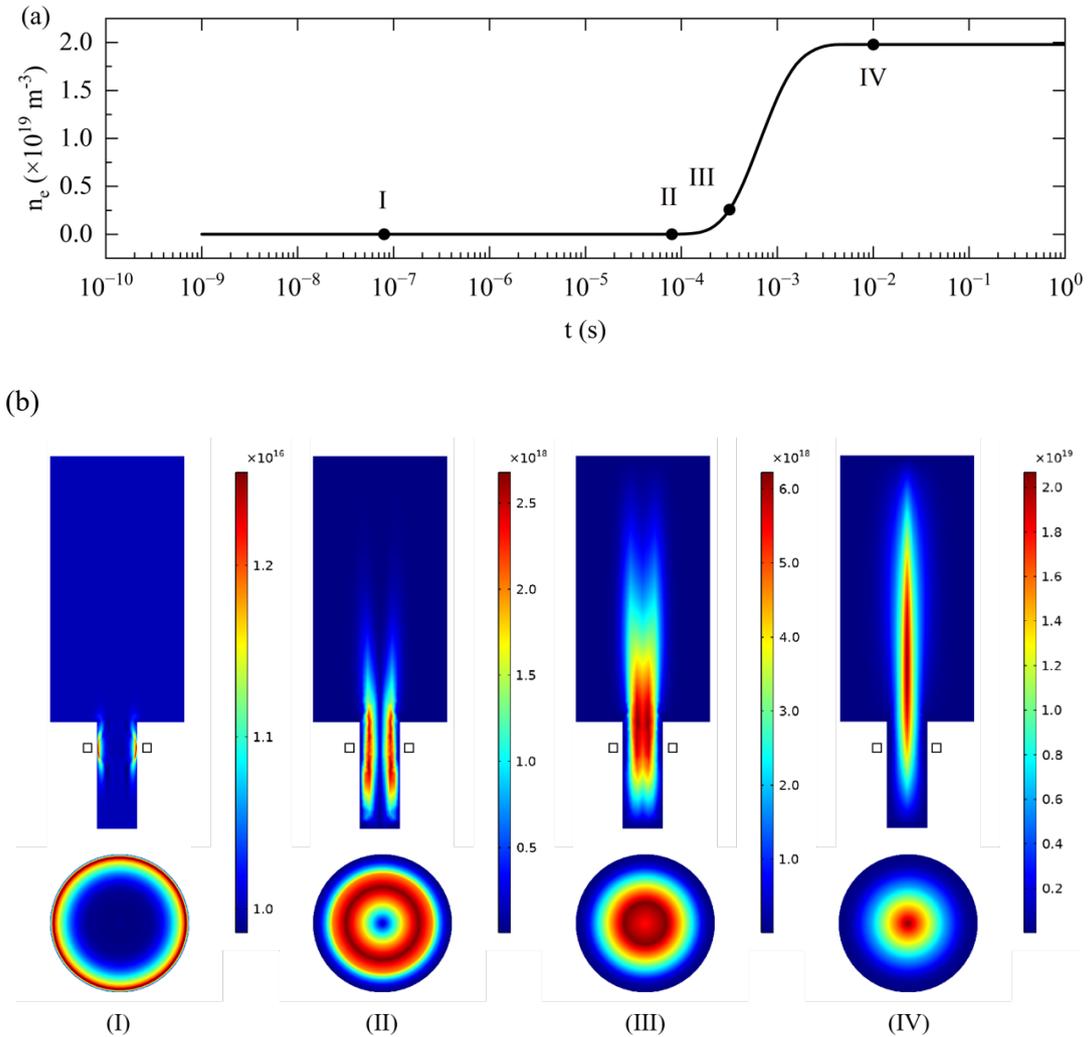

**Figure 4.** Temporal evolution of the electron density at r = 0 cm and z = 75 cm (a), and spatial distributions of electron density at four representative moments (b): 7.94×10$^{-8}$ s (I), 7.94×10$^{-5}$ s (II), 3.16×10$^{-4}$ s (III), and 0.01 s (IV).

- **Axial propagation and downstream filling**

Figure 5 provides complementary axial and radial cuts of the density evolution. Initially, the plasma column is localized near the antenna region. As the discharge strengthens, the high-density zone expands downstream, filling the central portion of the tube before gradually extending toward both ends. The radial profiles reveal that the fully developed discharge possesses a strongly peaked on-axis density with increasing radial steepness downstream—a signature of enhanced cross-field transport and reduced parallel losses in that region. The steady-state peak density reaches ~$2.5 \times 10^{19}$ m$^{-3}$, consistent with the density range typically observed in high-power helicon sources.

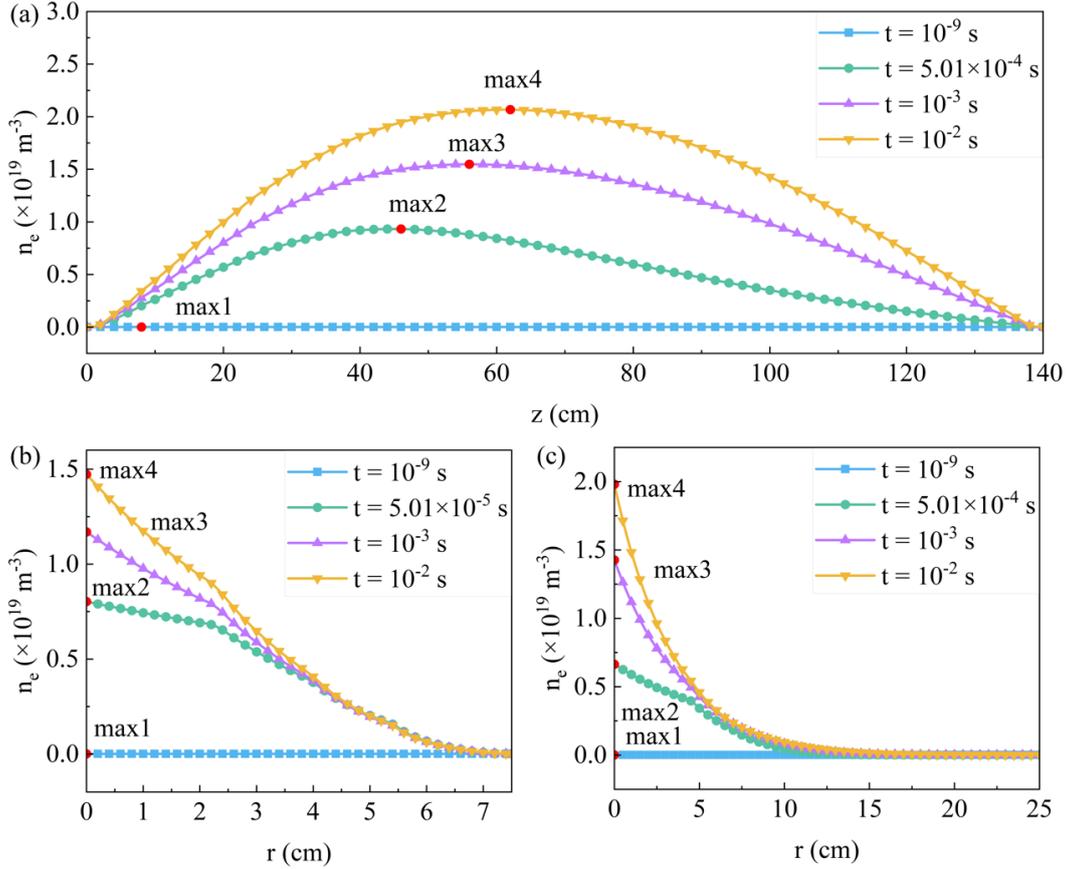

**Figure 5.** Axial and radial distributions of electron density: (a) axial distribution (r = 0 cm, z = 0-140 cm); (b) radial distribution at the antenna position (r = 0-7.5 cm, z = 30 cm); (c) radial distribution at the downstream probe position (r = 0-25 cm, z = 75 cm).

- **Temporal dynamics of electron heating: dual-peak evolution**

The electron temperature exhibits a markedly different temporal structure. Figure 6 shows that the temperature rises sharply immediately after breakdown, reaching a transient maximum of ~2.25 eV at ~$5 \times 10^{-7}$ s. This initial overshoot indicates a transient imbalance between electron heating and collisional/inelastic energy losses, after which the mean electron energy decreases as ionization and inelastic processes intensify. A secondary, broader temperature peak of ~3 eV occurs at ~$5 \times 10^{-6}$ s, coinciding with accelerated ionization and rapid density growth. This dual-peak behavior is consistent with Lieberman's approximate energy balance model [64], where the competition between heating and ionization energy loss governs the transient response. Spatially, the highest temperatures initially appear near the antenna and along the outer radius (Fig. 6(b)); the temperature profile then gradually flattens and relaxes to a steady value of ~1.25 eV.

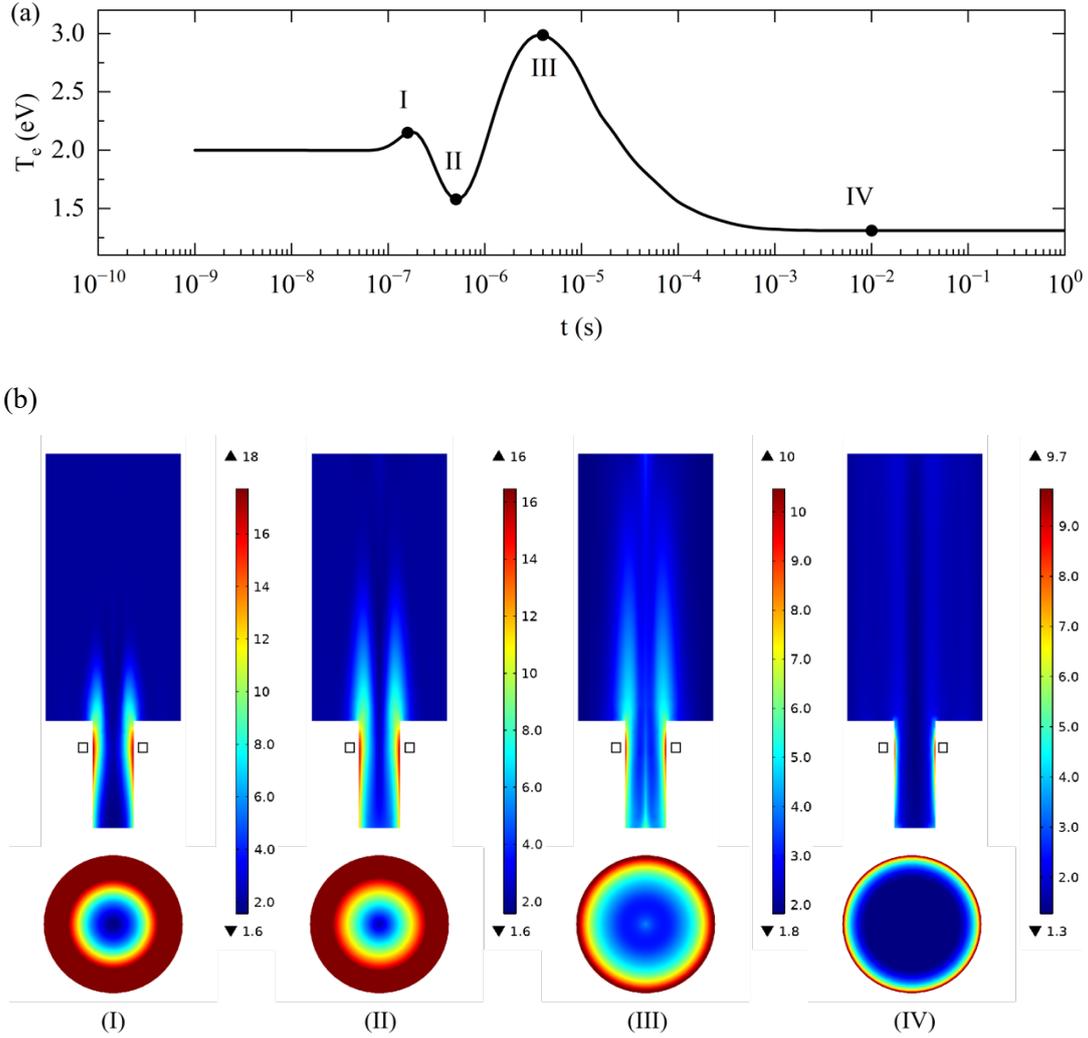

**Figure 6.** Temporal evolution of the electron temperature at r = 0 cm and z = 75 cm (a), and spatial distributions of electron temperature at four representative moments (b): $1.58\times10^{-7}$ s (I), $5.01\times10^{-7}$ s (II), $3.98\times10^{-6}$ s (III), and 0.01 s (IV).

- **Temperature structure along axis and radius**

Figure 7 shows that the early-time electron temperature can exceed 6 eV upstream, while remaining lower downstream. A strong radial temperature gradient develops near the antenna (z = 30 cm). Further downstream (z = 75 cm), a transient on-axis temperature peak forms due to the delayed penetration of wave energy and the rapid rise of the local density. As the discharge stabilizes, collisional cooling smooths the temperature gradients and the system relaxes toward its steady state.

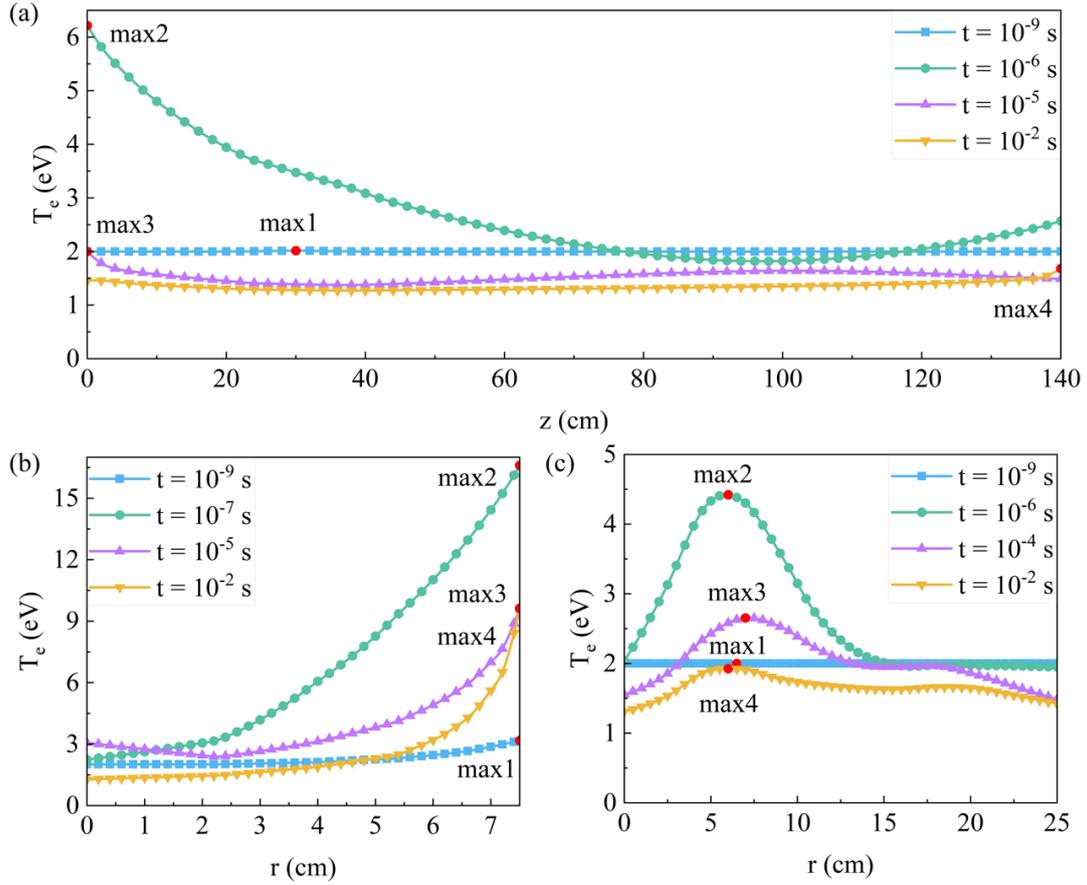

**Figure 7.** Axial and radial distributions of electron temperature: (a) axial distribution (r = 0 cm, z = 0-140 cm); (b) radial distribution at the antenna position (r = 0-7.5 cm, z = 30 cm); (c) radial distribution at the downstream probe position (r = 0-25 cm, z = 75 cm).

- **Physical interpretation and broader significance**

Taken together, Figures 4–7 reveal several universal features of helicon plasma ignition. First, we observe edge-dominated initial ionization. Further, we also see inward migration of the density peak, namely as conductivity increases, helicon-mode coupling strengthens and shifts power deposition toward the axis, leading to core formation. Then, it is very interesting to find dual-peak electron temperature evolution, i.e. a rapid overshoot followed by a slower relaxation reflects the redistribution of RF energy among a rapidly growing electron population. Moreover, one can see axial filling of the discharge chamber, which means that plasma develops first near the antenna and subsequently fills the diffusion chamber, a sequence consistent with time-resolved optical and probe diagnostics reported in helicon experiments. These dynamics are not merely device-specific: they illustrate the fundamental nonequilibrium physics governing RF plasma ignition in magnetized systems. The interplay between wave propagation, evolving dielectric response, and collisional energy transfer is central to helicon thrusters, industrial etching reactors, and helicon-based density control in magnetic fusion devices. The present model provides a predictive framework for optimizing the stability, uniformity, and efficiency of next-generation RF plasma sources.

## 3.2 Spatiotemporal wave field and power deposition

To further elucidate the dynamics of helicon plasma formation, we examine the spatiotemporal behavior of the RF wave fields and the associated power deposition in both axial and radial directions. Figures 8–10 present the evolution of the wave electric field, wave magnetic field, and power deposition, respectively.

- **Wave electric field evolution**

Figure 8 shows the evolution of the wave electric field, highlighting the transition from vacuum-like excitation to plasma-modified mode structures. At the earliest times (~$10^{-9}$ s), the RF electric field is concentrated near the antenna and rapidly intensifies to nearly 500 V/m as the first electrons are created. Near the antenna position (z = 30 cm), the strongest field appears close to the boundary (r ≈ 7.5 cm). Further downstream at z = 75 cm, the radial profiles evolve significantly as the plasma density increases. During the interval $10^{-9}$–$10^{-5}$ s, the plasma remains too weak to perturb the wave, and the field amplitude remains small. Once the plasma density rises to the $10^{18}$–$10^{19}$ m$^{-3}$ range (~$10^{-4}$ s), the increased conductivity alters the dielectric response. As the discharge approaches steady state (~$10^{-2}$ s), the radial electric field becomes increasingly edge-localized. The enhanced core conductivity raises the effective permittivity near the axis and reduces the wave penetration depth, causing most of the RF electric field—and consequently the power deposition—to migrate toward the plasma boundary.

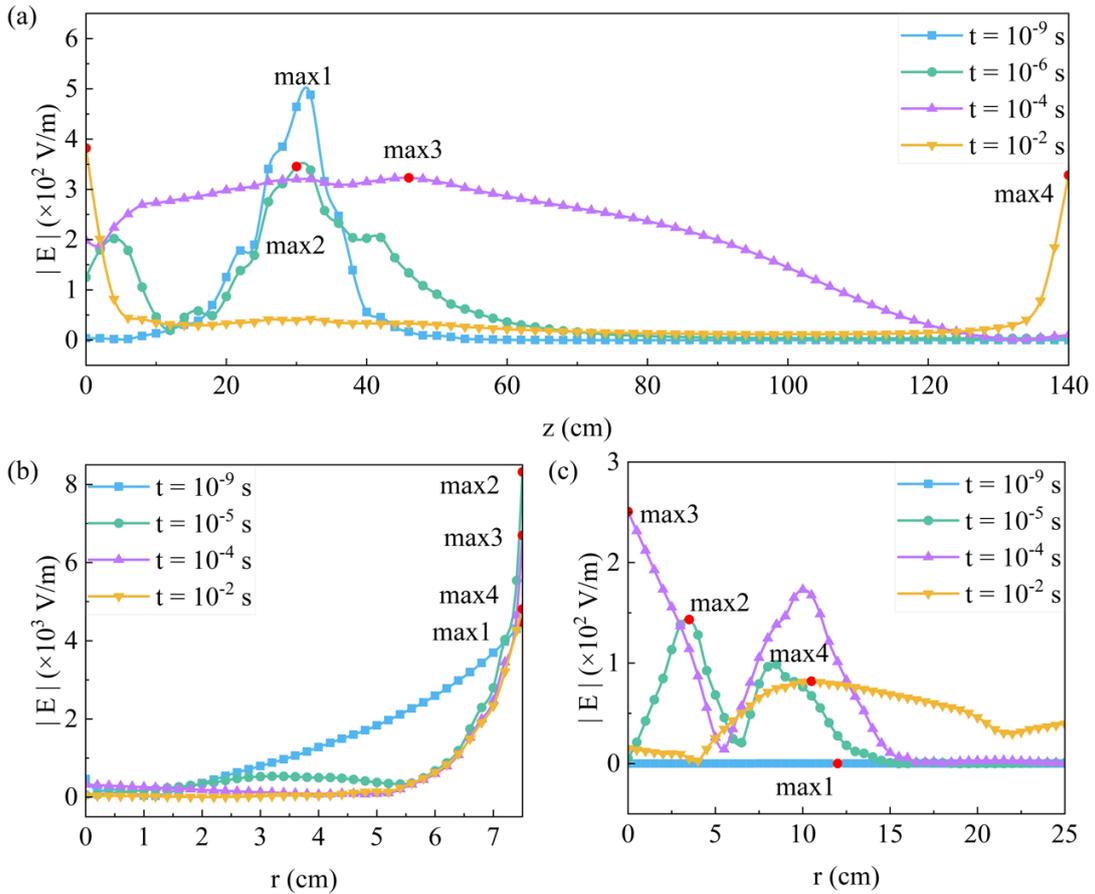

**Figure 8.** Axial and radial distributions of the wave electric field: (a) axial distribution at r = 0 cm; (b) radial distribution at z = 30 cm; (c) radial distribution at z = 75 cm.

- **Wave magnetic field evolution**

Figure 9 displays the corresponding spatiotemporal evolution of the RF magnetic field. At the earliest stage (~$10^{-9}$ s), the magnetic field follows the vacuum response of the antenna because the plasma density is extremely low. This results in a pronounced on-axis maximum and fast axial decay. Between $10^{-6}$ and $10^{-4}$ s, the emerging plasma alters the field structure. At $z = 30$ cm, the initially strong on-axis peak is gradually suppressed as the plasma becomes more conductive. In this regime, the plasma column acts as a conducting core, causing the RF magnetic flux to be displaced radially outward. Downstream at $z = 75$ cm, the field amplitude decreases to the milligauss level due to damping and geometric spreading during propagation. The radial profile becomes relatively flat, exhibiting only a modest enhancement near the edge.

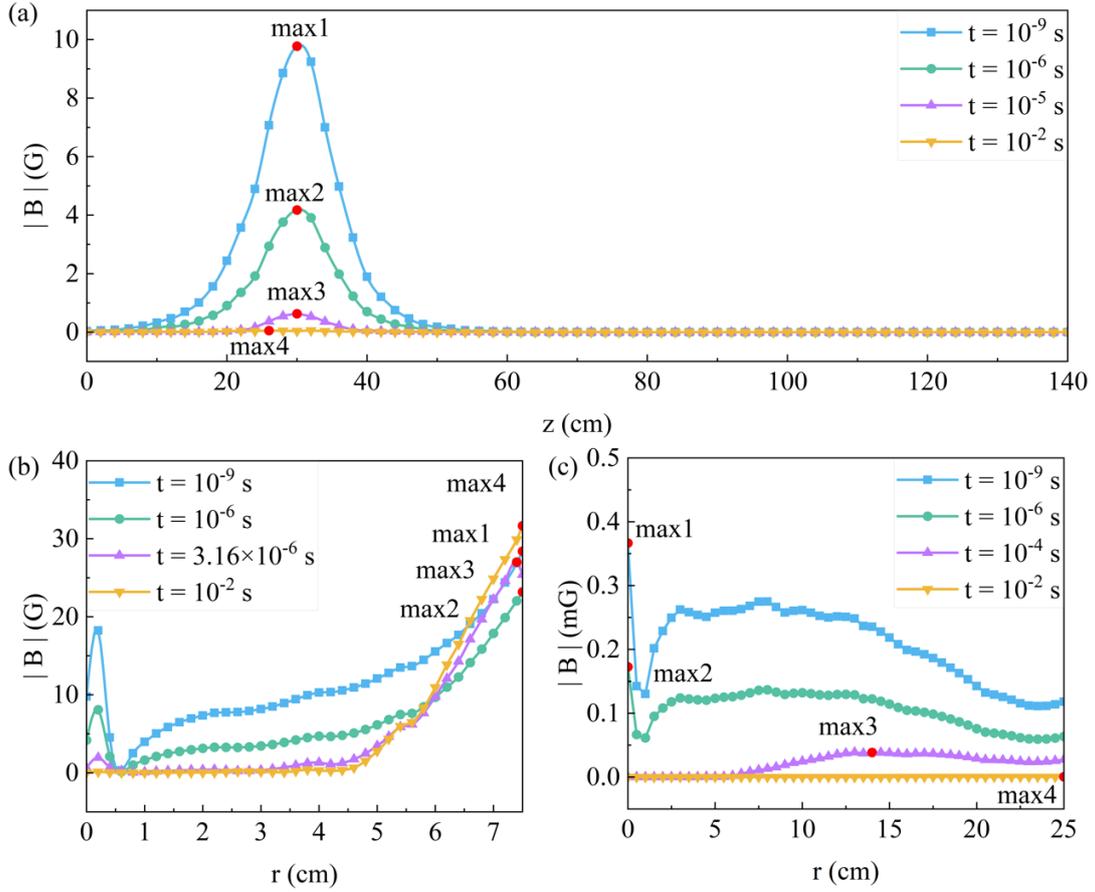

**Figure 9.** Axial and radial distributions of the wave magnetic field: (a) axial distribution at $r = 0$ cm; (b) radial distribution at $z = 30$ cm; (c) radial distribution at $z = 75$ cm.

- **Power deposition evolution**

Figure 10 presents the spatiotemporal evolution of RF power deposition, revealing how energy transfer mechanisms evolve as the plasma develops. Initially, power absorption is concentrated near the antenna at $z \approx 30$ cm, where the peak power density briefly exceeds $4 \times 10^4$ W/m$^3$ before rapidly diminishing. The corresponding radial profile shows a sharp absorption maximum near $r = 6$–7 cm, while the plasma core receives virtually no power in this early stage, which is a feature that is consistent with the off-axis power deposition near $z \approx 30$ cm observed in Ref. [35]. As ionization proceeds, the peak power density at $z = 30$ cm increases from ~$3 \times 10^6$ to ~$6 \times 10^6$ W/m$^3$, and it consistently remains in the outer radial region.

Further downstream at $z = 75$ cm, the radial distribution becomes much flatter and the overall power deposition decreases sharply, indicating substantial damping of wave energy during propagation. The center receives little power throughout the discharge, especially after the steady conductive core has formed.

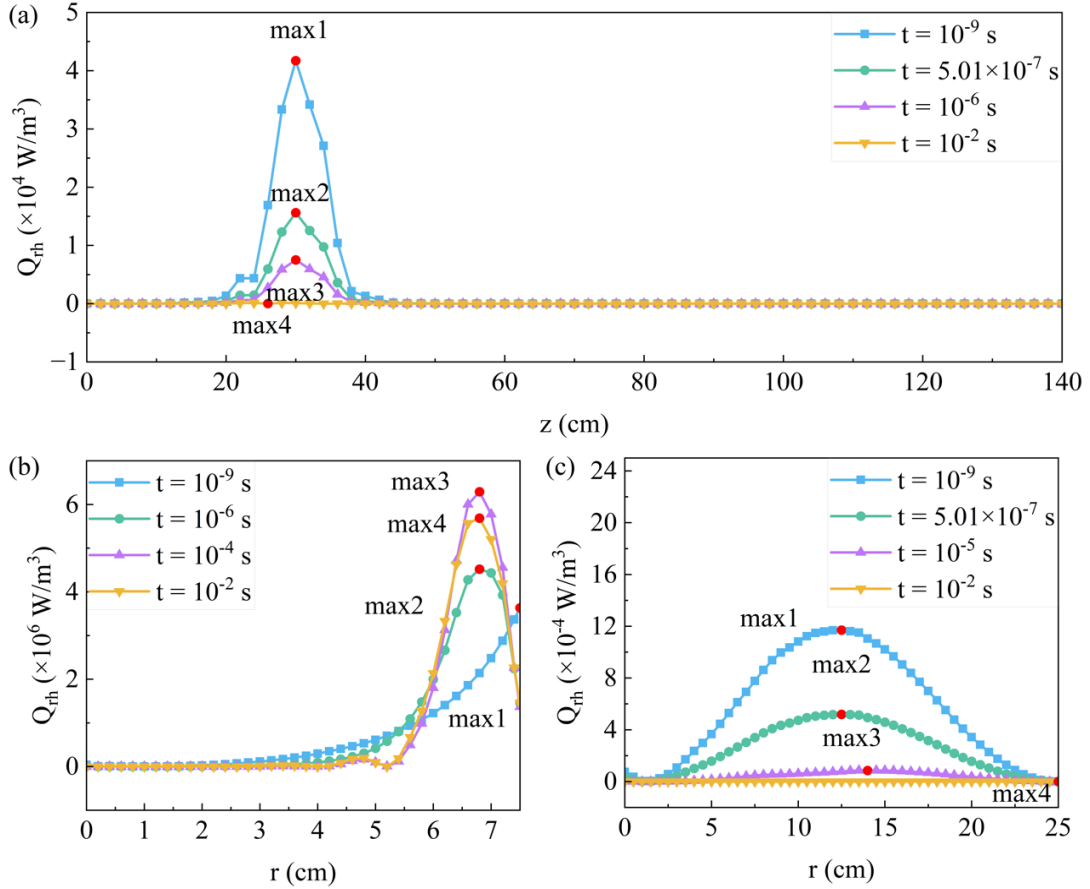

**Figure 10.** Axial and radial distributions of power deposition: (a) axial distribution at r = 0 cm; (b) radial distribution at z = 30 cm; (c) radial distribution at z = 75 cm.

Overall, Figures 4–10 collectively reveal a coherent physical picture of helicon discharge evolution: breakdown initiates at the periphery, where the RF wave fields are strongest; ionization fronts propagate inward and downstream, forming the core plasma column; electron temperature peaks in the outer region, reflecting peripheral heating pathways; electric fields, magnetic fields, and power deposition remain strongly edge localized, particularly in the steady state; helicon-mode contributions appear transiently during intermediate density buildup but become depleted as conductivity increases; edge-localized wave heating ultimately dominates the mature discharge and sustains the plasma. This integrated view highlights the essential role of wave–plasma feedback: as the plasma density rises, it reshapes the RF field topology, thereby redirecting where power is deposited and governing the spatial structure of the final steady-state discharge. These processes are fundamental not only to helicon sources but to a wider class of RF-driven plasmas where dielectric evolution and surface-wave phenomena control energy transport and sustainment.

### 3.3 Parameter-scan Study

Systematically varying key operating parameters provides deeper insight into the mechanisms that govern helicon plasma formation, wave–plasma coupling, and steady-state sustainment. Here, we examine how the neutral gas pressure, external magnetic field strength, antenna driving frequency, and input RF power influence the temporal evolution of plasma parameters. Figures 11–14 summarize the behavior of electron density, electron temperature, wave electric and magnetic fields, and RF power deposition at a representative location near the antenna ($r = 0$ cm, $z = 30$ cm).

- **Influence of neutral gas pressure**

Neutral pressure plays a central role in determining collision frequencies between electrons, ions, and neutrals, thereby shaping ionization dynamics, wave damping, and energy deposition. Figure 11 compares the temporal evolution of key plasma quantities for neutral pressures of 0.5, 2.5, 5.0, and 10.5 Pa.

For all pressures, the electron density rises slowly at first and then undergoes a sharp increase near $10^{-4}$ s, marking the transition from weak ionization to a self-sustained ionization regime. The steady-state density increases strongly with pressure. This behavior is consistent with the electron continuity relation [65]

$$\frac{dn_e}{dt} = K_{ion} n_g n_e - \frac{n_e u_B}{d_{eff}} - \alpha n_e^2 \quad (27)$$

where $K_{ion}$ is the ionization rate coefficient, $n_g$ is the neutral gas density, $u_B$ is the Bohm speed, $d_{eff}$ is the effective loss length, and $\alpha$ is the recombination coefficient. Before $10^{-4}$ s, the electron density is too low to produce an ionization rate comparable to diffusion and recombination losses, thereby the discharge remains in a slow-growth regime. When the accumulated electrons reach a critical level such that the ionization term $K_{ion} n_g n_e$ becomes comparable to or exceeds the loss term $n_e u_B / d_{eff}$, the discharge enters a self-reinforcing ionization phase and the electron density rises rapidly. Because the characteristic loss time governed by $u_B$ and $d_{eff}$ is primarily determined by RF power, magnetic configuration rather than neutral pressure, the ignition time of the rapid rise remains nearly the same for all pressure conditions. The electron temperature for all pressures exhibits the characteristic two-peak behavior described in Sec. 3.1: an initial overshoot governed by early electron heating, followed by a secondary maximum as ionization intensifies and energy redistributes. As pressure increases, the peak temperature monotonically decreases. Higher neutral density leads to more frequent elastic and inelastic collisions, enhancing energy dissipation and reducing the heating efficiency. Both the electric and magnetic field amplitudes show a non-monotonic dependence on pressure in the early stage (t ≲ $10^{-9}$ s). At low pressure, weak collisional damping allows strong initial fields. At intermediate pressures (2.5–5 Pa), enhanced collisional losses suppress the field amplitude. At the highest pressure, growing plasma density alters the dielectric response and shifts wave power radially outward, partially restoring the field amplitude. At later times, all wave fields decrease sharply as the plasma becomes conductive and absorbs RF energy. Higher pressure consistently produces lower steady-state RF field amplitudes due to stronger collisional damping and shorter penetration depth. Higher pressure leads to stronger collisional dissipation and more localized RF power deposition, causing the heating pattern to shift from volume-dominated absorption to a more surface-dominated pathway that may involve TG-like mode contributions. This trend is consistent with earlier studies on collisional wave damping in helicon devices [66].

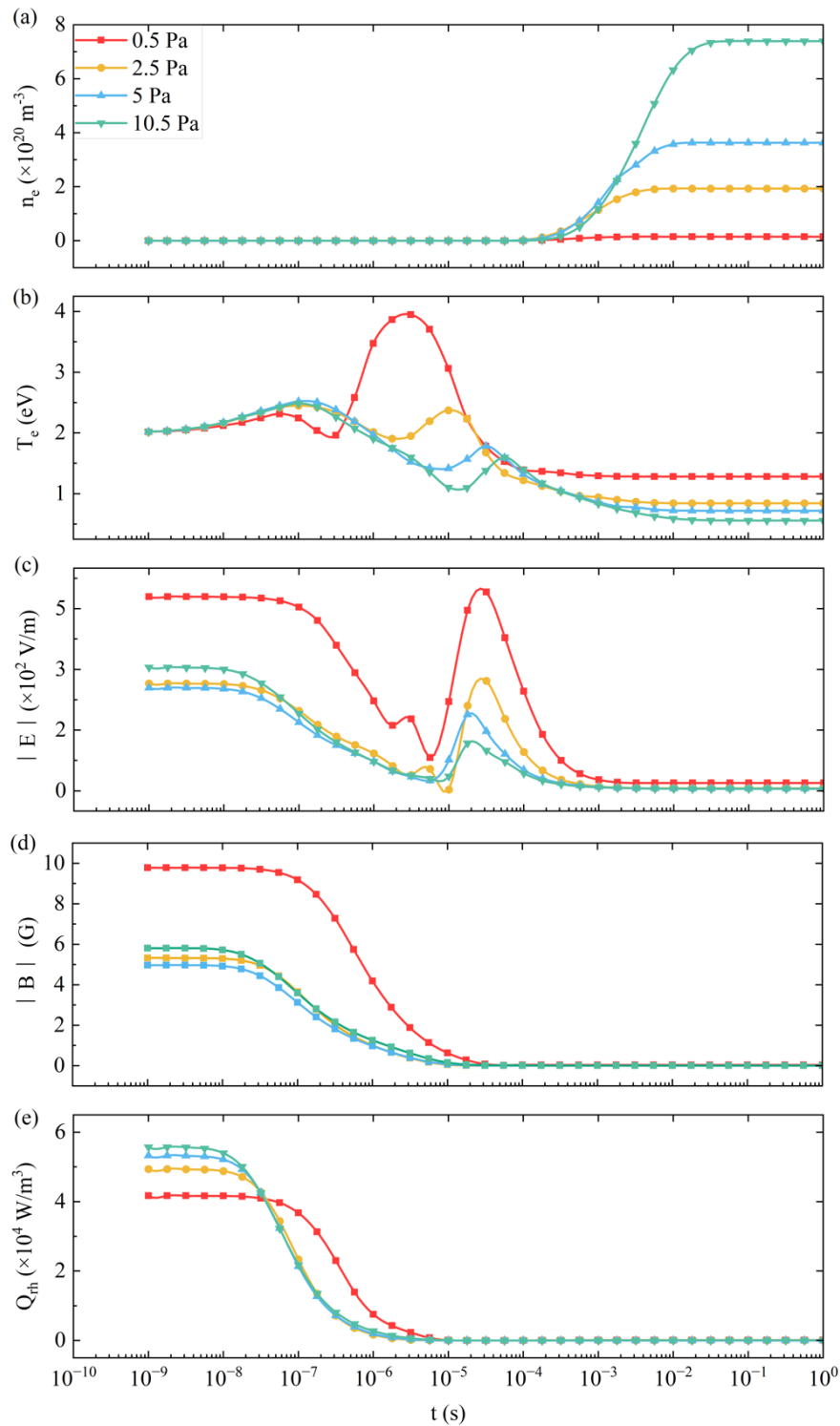

**Figure 11.** Temporal evolution of plasma parameters at the antenna position (r = 0 cm, z = 30 cm) for different neutral gas pressures (0.5, 2.5, 5.0, and 10.5 Pa): (a) electron density, (b) electron temperature, (c) wave electric field amplitude, (d) wave magnetic field amplitude, (e) RF power deposition.

In summary, increasing neutral pressure accelerates ionization and raises steady-state density, suppresses electron temperature peaks through enhanced collisional cooling, strengthens wave damping and reduces field penetration, and shifts power deposition from core (volumetric) heating toward edge (surface) heating. Neutral pressure therefore governs not only the steady-state plasma properties but also the early ignition dynamics and wave-energy deposition pathways.

- **Influence of external magnetic field**

The external magnetic field governs both electron confinement and the accessibility of helicon, thereby affecting ionization, power coupling, and wave propagation. Figure 12 shows the evolution of key quantities for field strengths of 500, 1000, 1500, and 2000 G.

For all field strengths, the electron density rises rapidly and then stabilizes. Higher magnetic fields significantly increase the steady-state density. Improved confinement increases the electron residence time and enhances wave–plasma coupling, enabling more efficient ionization. This trend is consistent with observations in the MPS-LD and other helicon devices [67]. At 500 G, a transient density overshoot appears before relaxing to a lower steady value. This two-stage evolution reflects the "ion-pumping" effect: rapid early ionization temporarily depletes neutrals near the axis, slowing subsequent ion production until neutrals diffuse back inward. This mechanism agrees with the neutral-depletion model of Cho [65]. Electron temperature evolution is remarkably insensitive to magnetic field strength. Peak temperatures of ~3.5–4 eV occur near $10^{-6}$ s, followed by a decline to ~1.5–2 eV. Because temperature is governed primarily by the balance between RF heating and collisional losses—both only weakly dependent on $B$—the magnetic field influences density much more strongly than temperature. Prior to plasma formation, the wave electric field is nearly identical for all $B$ since the medium is essentially vacuum. Once ionization accelerates, the field amplitude collapses rapidly due to increased conductivity. The steady-state electric field is almost identical for all field strengths, indicating that the antenna current—rather than $B$—sets the ultimate field levels after the plasma becomes fully ionized. Wave magnetic field and power deposition behave similarly and show minimal dependence on $B$ within the tested range, so they are omitted for brevity.

Therefore, higher external magnetic field significantly enhances electron density through improved confinement, produces minimal change in electron temperature and wave fields, and does not strongly affect steady-state power deposition. Most of the benefit of strong applied fields stems from enhanced helicon-wave propagation and confinement rather than direct modification of energy balance.

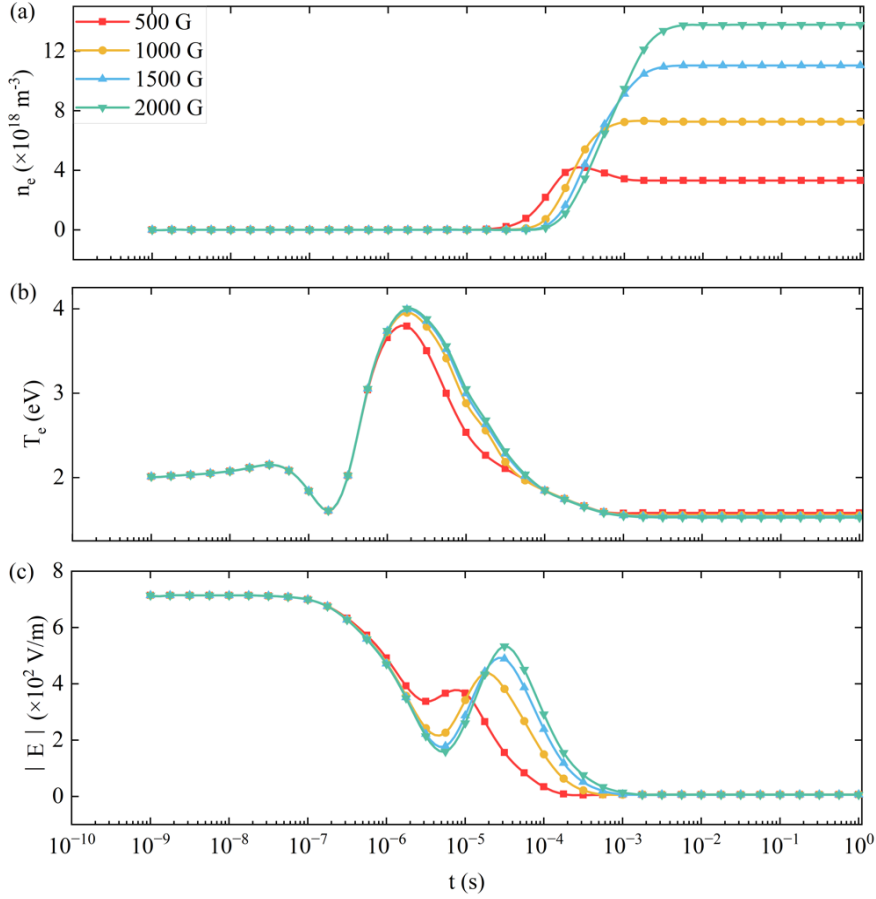

**Figure 12.** Temporal evolution of plasma parameters at the antenna position (r = 0 cm, z = 30 cm) for different external magnetic field strengths (500, 1000, 1500, and 2000 G): (a) electron density, (b) electron temperature, (c) wave electric field amplitude.

- **Influence of antenna driving frequency**

The antenna driving frequency influences wave propagation, the radial distribution of power deposition, and the overall discharge efficiency. Figure 13 compares behavior at 6.78, 13.56, 27.12, and 40.68 MHz.

All frequencies exhibit the characteristic slow-rise followed by rapid density growth near $10^{-4}$ s. However, the steady-state density decreases monotonically with increasing frequency. At low frequencies, helicon-mode coupling is stronger and power deposition favors the central region, supporting efficient ionization. As frequency increases, wave energy shifts outward toward more edge-localized modes, reducing core heating and lowering density. Temperature evolution shows the same rapid-rise/peak/relaxation pattern discussed earlier. The peak temperature (~4 eV) and steady value (~1.5 eV) vary little with frequency. This insensitivity reflects the fact that collisional processes—not wave frequency—dominate electron energy balance in the parameter ranges considered.

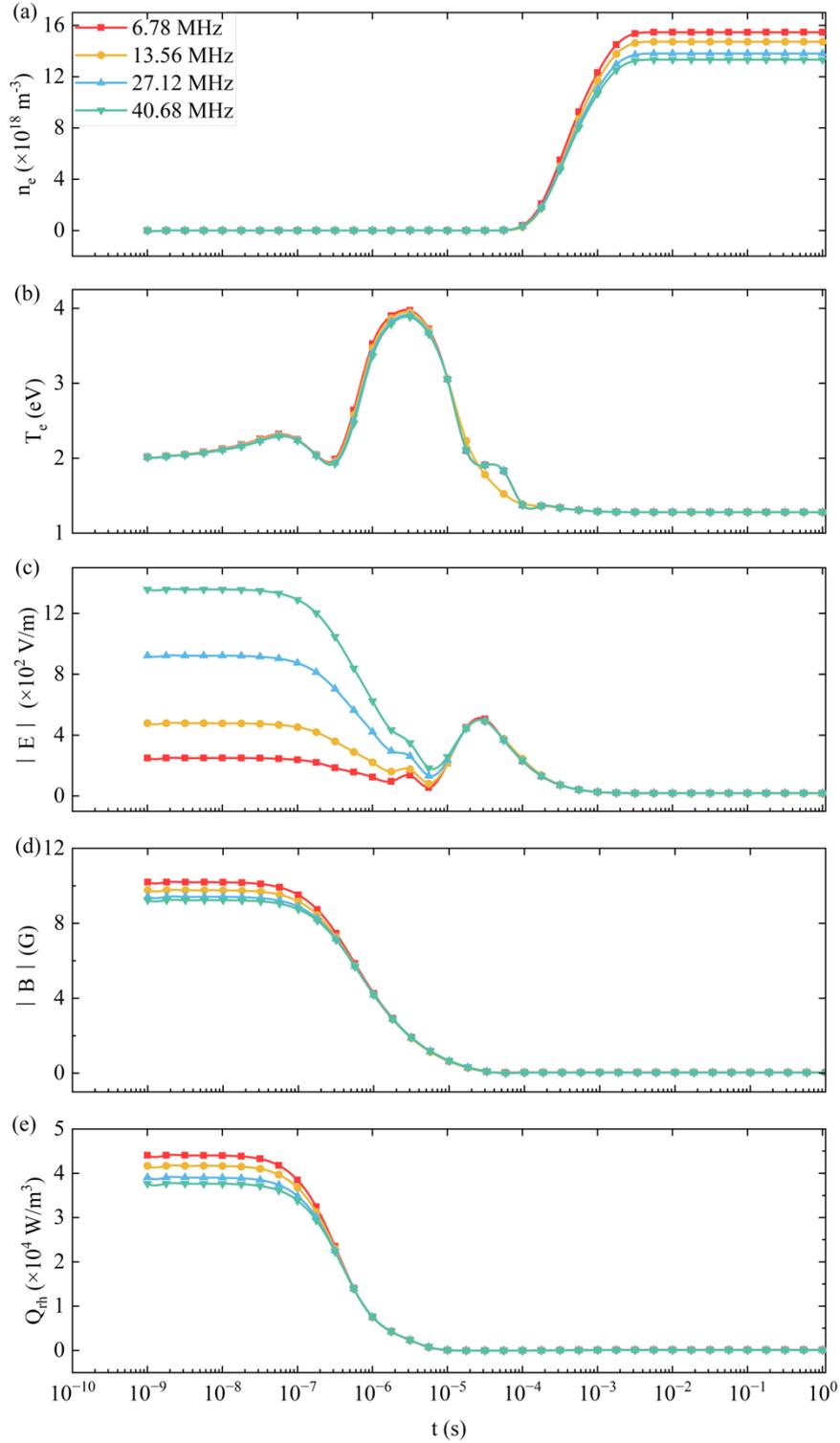

**Figure 13.** Temporal evolution of plasma parameters at the antenna position (r = 0 cm, z = 30 cm) for different antenna driving frequencies (6.78, 13.56, 27.12, and 40.68 MHz): (a) electron density, (b) electron temperature, (c) wave electric field amplitude, (d) wave magnetic field amplitude, (e) RF power deposition.

The initial wave electric field amplitude increases strongly with frequency, reaching ~1400 V/m at 40.68 MHz but only tens of V/m at 6.78 MHz. Higher-frequency waves are less strongly shielded and retain larger initial amplitudes. Wave magnetic fields show the opposite trend: initial amplitudes decrease modestly with increasing frequency. Both electric and magnetic fields decay rapidly as plasma forms, converging toward similar steady values. Power deposition follows behavior consistent with the effective conductivity, i.e. Eq. (21). At low frequency, the conductivity is large and Ohmic heating is efficient. At high frequency, the conductivity decreases as $\omega^{-2}$, suppressing energy absorption and reducing ionization efficiency.

Thus, higher driving frequencies reduce steady-state electron density by shifting wave energy outward, modestly affect wave magnetic fields and temperature evolution, and weaken Ohmic heating through reduced effective conductivity. Low-frequency excitation thus offers better core heating and ionization efficiency in helicon devices.

- **Influence of input RF power**

RF power determines the total available energy for electron heating, ionization, and wave excitation. Figure 14 presents results for input powers of 350, 450, 1000, and 1350 W.

Electron density increases rapidly between $10^{-4}$ and $10^{-3}$ s for all power levels. Higher input power yields higher density at all times due to stronger ionization and enhanced electron accumulation. At 1350 W, the central density exceeds $8\times10^{18}$ m$^{-3}$—more than triple the value at 350 W. Temperature evolution shows a sharp rise followed by a peak near $\sim10^{-5}$ s. With increasing power, the transient temperature peak becomes noticeably higher, reflecting more intense early electron heating. The steady-state temperature remains ~1.5 eV across all powers, as collisional cooling eventually balances the higher power input. Wave electric fields remain large prior to plasma formation but collapse rapidly once the plasma becomes conducting. Higher RF power accelerates this collapse, indicating faster growth of conductivity and stronger wave absorption. The magnetic field amplitude initially increases with power—from ~7 G at 350 W to ~14 G at 1350 W—reflecting stronger antenna excitation. However, the steady-state magnetic-field amplitude decreases with increasing power. Higher density and conductivity produce stronger electromagnetic shielding, limiting wave penetration and reducing the sustained magnetic field in the plasma interior. Power deposition shows the same rapid decline after ignition but reaches substantially higher peak values at higher input power, with $>3\times10^4$ W/m$^3$ at 1350 W. This reflects more efficient power coupling and faster ionization at elevated power.

Hence, increasing RF power accelerates ionization and raises steady-state density, enhances early-time electron heating but leaves final temperature nearly unchanged, strengthens electromagnetic shielding at high densities, reducing steady-state wave fields, and substantially increases peak and total power deposition. High RF power improves discharge efficiency but also increases wave damping and surface-localized heating in the steady state.

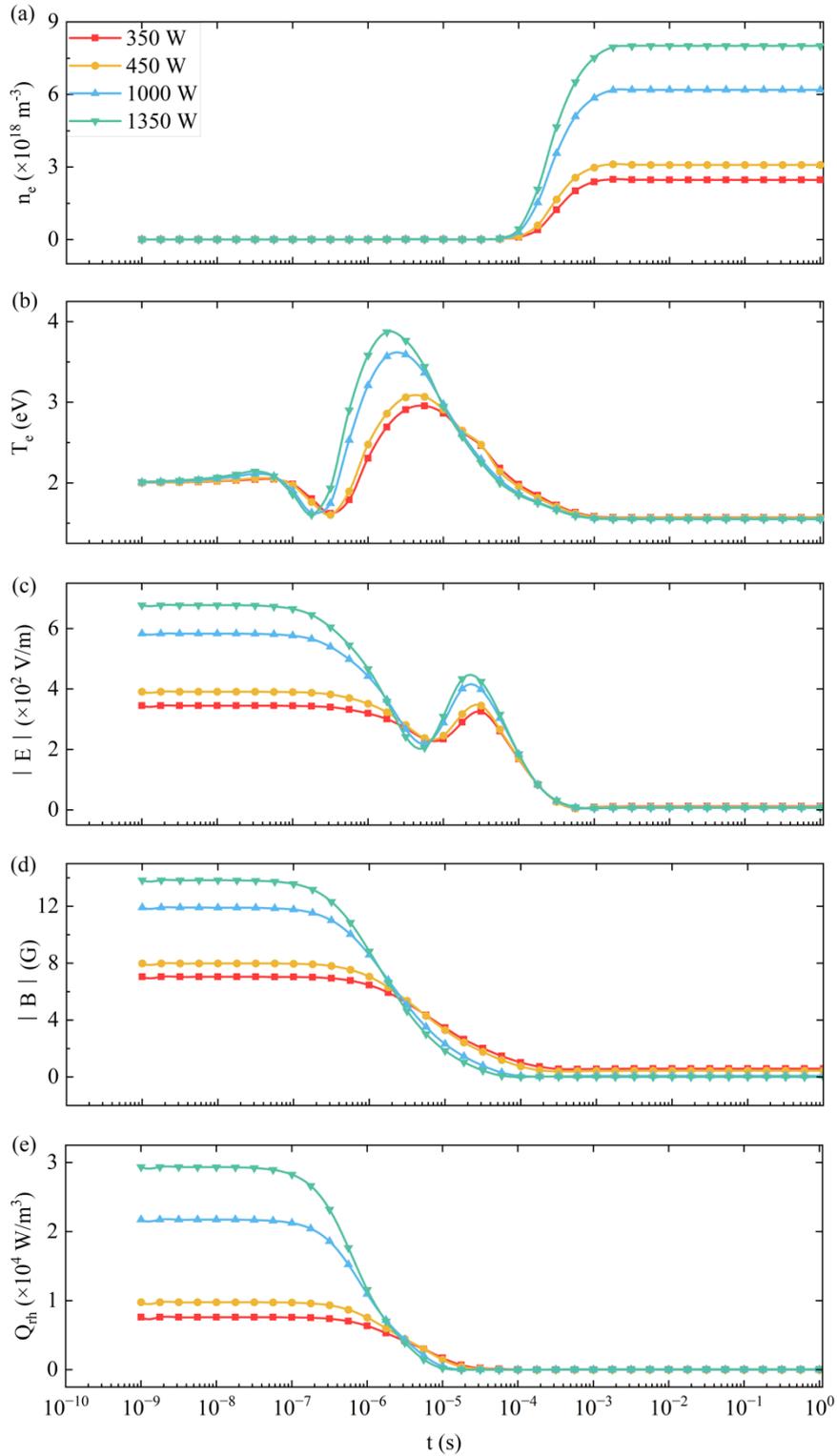

**Figure 14.** Temporal evolution of plasma parameters at the antenna position (r = 0 cm, z = 30 cm) for different input RF powers (350, 450, 1000, and 1350 W): (a) electron density, (b) electron temperature, (c) wave electric field amplitude, (d) wave magnetic field amplitude, (e) RF power deposition.

## 4. Conclusion

In summary, we have developed a fully self-consistent multiphysics framework that captures the complete spatiotemporal evolution of helicon plasma ignition, enabling the transient ionization dynamics and wave–plasma interaction processes to be resolved with unprecedented clarity. The model couples electron drift–diffusion transport, heavy-particle kinetics, electromagnetic wave propagation, and key chemical processes, and reproduces the experimentally observed parameter trends of the PPT device across magnetic field, pressure, frequency, and power ranges. The simulations uncover a distinct transient ionization stage, during which the electron density exhibits a rapid jump at $\sim 10^{-4}$ s while the electron temperature undergoes a dual-peaked evolution before relaxing toward steady state. Through analysis of the RF field topology and power deposition, we infer a possible role of nonlinear coupling between Trivelpiece–Gould and helicon modes in driving the observed transient behavior. Gas pressure, magnetic field strength, driving frequency, and RF power are shown to regulate this mode-coupling process in different ways: pressure and magnetic field primarily enhance steady-state density and suppress temperature overshoot, frequency determines the stability of RF power deposition, and RF power controls the rate of density growth. The quantitative agreement between simulated trends and available measurements indicates that the model provides a realistic description of the discharge, and it offers clear guidance for future time-resolved diagnostics—such as Langmuir probe scans, optical emission measurements, and RF field probes—to directly probe the predicted density jump and temperature overshoot. Remaining discrepancies in the earliest stage likely reflect the influence of probe perturbation, neutral nonuniformity, and antenna losses, which can be incorporated into future model refinements. More broadly, the parameter-dependent mechanisms identified here—particularly the interplay of wave propagation, collisional energy transfer, and transient ionization—establish a unified physical picture of helicon discharge ignition. These insights are directly applicable to the optimization of high-power helicon thrusters, industrial RF plasma sources, and helicon-based density control schemes in magnetic fusion devices. Future developments will extend the present model to include electron convection, neutral temperature gradients, and three-dimensional antenna geometries, enabling predictive capability for a wider class of helicon plasma systems.

**Acknowledgements**

Discussions with Rod Boswell are appreciated. This work is supported by the National Natural Science Foundation of China (Nos. 92271113, 12411540222, 12481540165), the Fundamental Research Funds for Central Universities (No. 2022CDJQY-003), the Chongqing Entrepreneurship and Innovation Support Program for Overseas Returnees (No. CX2022004), the Natural Science Foundation Project of Chongqing (No. CSTB2025NSCQ-GPX0725), and the ENN's Hydrogen-Boron Fusion Research Fund (No. 2025ENNHB01-011).

# References

[1]  Isayama S., Shinohara S. & Hada T. Review of helicon high-density plasma: Production mechanism and plasma/wave characteristics. *Plasma Fusion Res.* **13**, 1101014 (2018).

[2]  Boswell R. W. Plasma production using a standing helicon wave. *Phys. Lett. A* **33**, 457–458 (1970).

[3]  Shinohara S. Helicon high-density plasma sources: Physics and applications. *Adv. Phys.: X* **3**, 1420424 (2018).

[4]  Chen F. F. Plasma ionization by helicon waves. *Plasma Phys. Control. Fusion* **33**, 339–364 (1991).

[5]  Chen F. F. & Torreblanca H. Large-area helicon plasma source with permanent magnets. *Plasma Phys. Control. Fusion* **49**, A81–A88 (2007).

[6]  Boswell R. W. Very efficient plasma generation by whistler waves near the lower hybrid frequency. *Plasma Phys. Control. Fusion* **26**, 1147–1162 (1984).

[7]  Boswell R. W. & Henry D. Pulsed high rate plasma etching with variable $Si/SiO_2$ selectivity and variable Si etch profiles. *Appl. Phys. Lett.* **47**, 1095–1097 (1985).

[8]  Takahashi K., Motomura T., Ando A. et al. Transport of a helicon plasma by a convergent magnetic field for high speed and compact plasma etching. *J. Phys. D: Appl. Phys.* **47**, 425201 (2014).

[9]  Chen F. F. Helicon discharges and sources: a review. *Plasma Sources Sci. Technol.* **24**, 014001 (2015).

[10] Komori A., Shoji T., Miyamoto K., Kawai J. & Kawai Y. Helicon waves and efficient plasma production. *Phys. Fluids B* **3**, 893–898 (1991).

[11] Carson J. & Sedwick R. Dynamics of a centrifugally confined fusion plasma in space propulsion applications. *J. Propul. Power* **41**, 313–319 (2025).

[12] Batishchev O. Mini-helicon plasma thruster. *IEEE Trans. Plasma Sci.* **37**, 1563–1571 (2009).

[13] Huang T., Jin C., Yu J. et al. One-step synthesis of silicon oxynitride films using a steady-state and high-flux helicon-wave excited nitrogen plasma. *Plasma Chem. Plasma Process.* **37**, 1237–1247 (2017).

[14] Díaz F. R. C. The VASIMR rocket. *Sci. Am.* **283**(5), 90–97 (2000).

[15] Ahedo E. & Merino M. Two-dimensional supersonic plasma acceleration in a magnetic nozzle. *Phys. Plasmas* **17**, 073501 (2010).

[16] Takahashi K. Magnetic nozzle radiofrequency plasma thruster approaching twenty percent thruster efficiency. *Sci. Rep.* **11**, 2768 (2021).

[17] Little J. M. & Choueiri E. Y. Thrust and efficiency model for electron-driven magnetic nozzles. *Phys. Plasmas* **20**, 103501 (2013).

[18] Charles C. Plasmas for spacecraft propulsion. *J. Phys. D: Appl. Phys.* **42**, 163001 (2009).

[19] Van Compernolle B., Brookman M. W., Moeller C. P. et al. The high-power helicon program at DIII-D: gearing up for first experiments. *Nucl. Fusion* **61**, 116034 (2021).

[20] Chen F. F. A compact permanent-magnet helicon thruster. *IEEE Trans. Plasma Sci.* **43**, 195–197 (2015).

[21] Takahashi K. & Ando A. Enhancement of axial momentum lost to the radial wall by the upstream magnetic field in a helicon source. *Plasma Phys. Control. Fusion* **59**, 054007 (2017).

[22] Bathgate S. N., Bilek M. M. & McKenzie D. R. Electrodeless plasma thrusters for spacecraft: a review. *Plasma Sci. Technol.* **19**, 083001 (2017).

[23] Takahashi K., Sugawara T. & Ando A. Spatially- and vector-resolved momentum flux lost to a wall in a magnetic nozzle rf plasma thruster. *Sci. Rep.* **10**, 1061 (2020).

[24] Shinohara S., Kuwahara D., Furukawa T., Nishimura S., Yamase T., Ishigami Y., Horita H., Igarashi A. & Nishimoto S. Development of featured high-density helicon sources and their application to electrodeless plasma thruster. *Plasma Phys. Control. Fusion* **61**, 014017 (2019).

[25] Wang S. J., Wi H. H., Kim H. J., Kim J., Jeong J. H. & Kwak J. G. Helicon wave coupling in KSTAR plasmas for off-axis current drive in high electron pressure plasmas. *Nucl. Fusion* **57**, 046010 (2017).

[26] Pinsker R. I., Prater R., Moeller C. P., deGrassie J. S., Petty C. C., Porkolab M., Anderson J. P., Garofalo A. M., Lau C., Nagy A. et al. Experiments on helicons in DIII-D—investigation of the physics of a reactor-relevant non-inductive current drive technology. *Nucl. Fusion* **58**, 106007 (2018).

[27] Zi D., Cheng Z., Sun S., & et al. Simulation of helicon wave current drive on the Experimental Advanced Superconducting Tokamak under high poloidal beta scenarios. *Fusion Engineering and Design* **215**, 114968 (2025).

[28] Chang L., Breizman B. N. & Hole M. J. Gap eigenmode of radially localized helicon waves in a periodic structure. *Plasma Phys. Control. Fusion* **55**, 025003 (2013).

[29] Zhang Y., Chang L. & Xu G. Wave penetration and power deposition of helicon current drive for magnetic fusion plasma with sharp edge gradient. *Plasma Phys. Control. Fusion* **67**, 065022 (2025).

[30] Islam M. S., Lore J. D., Lau C., & et al. Analysing the effects of heating and gas puffing in Proto-MPEX helicon and auxiliary heated plasmas. *Plasma Physics and Controlled Fusion* **65**, 095020 (2023).

[31] Pinsker R. I., Prater R., Moeller C. P. et al. Experiments on helicons in DIII-D—investigation of the physics of a reactor-relevant non-inductive current drive technology. *Nucl. Fusion* **58**, 106007 (2018).

[32] Sun M., Xu X., Wang C., et al. Effect of antenna helicity on discharge characteristics of helicon plasma under a divergent magnetic field. *Plasma Science and Technology* **26**, 064006 (2024).

[33] C Wang, L Chang, L-F Lu, S Shinohara, Z-D Zeng, I Zadiriev, E Kralkina, Z Li, S-J Zhang, Z-C Kan, Y Tao & D-Z Li. Spatial and temporal evolutions of blue-core helicon discharge driven by a planar antenna with concentric rings. *Phys. Plasmas* **32**, 113509 (2025).

[34] Chang L., Zhang S. J., Wu J. T. et al. Transition of blue-core helicon discharge. *arXiv preprint* **arXiv:**2508.19662 (2025).

[35] Y.-J. Chang, L. Chang, X.-G. Yuan, X. Yang, Q. Xu, Y. Wang, G.-J. Niu, H.-S. Zhou & G.-N. Luo. Numerical study on the temporal evolution of a helicon discharge. *IEEE Trans. Plasma Sci.* **49**, 3733–3744 (2021).

[36] Clarenbach B., Lorenz B., Krämer M. & Sadeghi N. Time-dependent gas density and temperature measurements in pulsed helicon discharges in argon. *Plasma Sources Sci. Technol.* **12**, 345–353 (2003).

[37] Ziemba T., Euripides P., Slough J., Winglee R., Giersch L., Carscadden J., Schnackenberg T. & Isley S. Plasma characteristics of a high power helicon discharge. *Plasma Sources Sci. Technol.* **15**, 517–525 (2006).

[38] Takahashi K., Takao Y. & Ando A. Neutral-depletion-induced axially asymmetric density in a helicon source and imparted thrust. *Appl. Phys. Lett.* **108**, 074103 (2016).

[39] Conway G. D., Perry A. J. & Boswell R. W. Evolution of ion and electron energy distributions in pulsed helicon plasma discharges. *Plasma Sources Sci. Technol.* **7**, 337–344 (1998).

[40] Biloiu I. A., Scime E. E. & Biloiu C. Time evolution of fast ions created in an expanding helicon plasma. In *2008 IEEE 35th International Conference on Plasma Science* 1–1 (IEEE,

2008).

[41] Biloiu C., Sun X., Choueiri E., Doss F., Scime E., Heard J., Spektor R. & Ventura D. Evolution of the parallel and perpendicular ion velocity distribution functions in pulsed helicon plasma sources obtained by time resolved laser induced fluorescence. *Plasma Sources Sci. Technol.* **14**, 766–776 (2005).

[42] Biloiu I. A. & Scime E. E. Temporal evolution of bimodal argon-ion velocity distribution in an expanding helicon plasma. *IEEE Trans. Plasma Sci.* **36**, 1376–1377 (2008).

[43] Stollberg C., Guittienne P., Karimov R. et al. First Thomson scattering results from AWAKE's helicon plasma source. *Plasma Phys. Control. Fusion* **66**, 115011 (2024).

[44] Mingyang Wu, Chijie Xiao, Yue Liu et al. Effects of magnetic field on electron power absorption in helicon fluid simulation. *Plasma Sci. Technol.* **23**, 085002 (2021).

[45] Bose D., Govindan T. R. & Meyyappan M. Modeling of a helicon plasma source. *IEEE Trans. Plasma Sci.* **31**, 464–470 (2003).

[46] I Isayama S., Shinohara S., Hada T. et al. Spatio-temporal behavior of density jumps and the effect of neutral depletion in high-density helicon plasma. *Phys. Plasmas* **26**, 053513 (2019).

[47] Y. Chang, L. Chang, X. Yuan et al. Numerical study on the temporal evolution of a helicon discharge. *IEEE Trans. Plasma Sci.* **49**, 3733–3744 (2021).

[48] Xiong Yang, Mousen Cheng, Dawei Guo et al. Characteristics of temporal evolution of particle density and electron temperature in helicon discharge. *Plasma Sci. Technol.* **19**, 105402 (2017).

[49] Afsharmanesh M. & Habibi M. A simulation study of the factors affecting the collisional power dissipation in a helicon plasma. *IEEE Trans. Plasma Sci.* **45**, 2272–2278 (2017).

[50] Mouzouris Y. & Scharer J. E. Wave propagation and absorption simulations for helicon sources. *Phys. Plasmas* **5**, 4253–4261 (1998).

[51] Boswell R. W. Dependence of helicon wave radial structure on electron inertia. *Aust. J. Phys.* **25**, 403–408 (1972).

[52] Tianchao Xu. Experimental study of turbulent transport on the PPT device and a novel magnetic field diagnostic method of Field-reversed Configuration (FRC). Ph.D. thesis, Peking University (2020).

[53] Mingyang Wu, Chijie Xiao, Yue Liu et al. Effects of magnetic field on electron power absorption in helicon fluid simulation. *Plasma Sci. Technol*. **23**, 085002 (2021).

[54] Chijie Xiao, Yang Xiong, Yue Liu et al. Plasma rotation in the Peking University Plasma Test device. *Rev. Sci. Instrum.* **87**, 11E107 (2016).

[55] Lymberopoulos D. P. & Economou D. J. Two-dimensional self-consistent radio frequency plasma simulations relevant to the gaseous electronics conference RF reference cell. J. Res. *Natl. Inst. Stan.* **100**, 473–495 (1995).

[56] Sternberg N., Godyak V. & Hoffman D. Magnetic field effects on gas discharge plasmas. *Phys. Plasmas* **13**, 063511 (2006).

[57] Hagelaar G. J. M. & Pitchford L. C. Solving the Boltzmann equation to obtain electron transport coefficients and rate coefficients for fluid models. *Plasma Sources Sci. Technol.* **14**, 722–733 (2005).

[58] Lymberopoulos D. P. & Economou D. J. Two-dimensional self-consistent radio frequency plasma simulations relevant to the gaseous electronics conference RF reference cell. *J. Res. Natl. Inst. Stan.* **100**, 473–495 (1995).

[59] Sternberg N., Godyak V. & Hoffman D. Magnetic field effects on gas discharge plasmas. *Phys. Plasmas* **13,** 063511 (2006).

[60] Clarenbach B., Krämer M. & Lorenz B. Spectroscopic investigations of electron heating in a high-density helicon discharge. *J. Phys. D: Appl. Phys.* **40**, 5117–5124 (2007).

[61] Hagelaar G. J. M. & Pitchford L. C. Solving the Boltzmann equation to obtain electron transport coefficients and rate coefficients for fluid models. *Plasma Sources Sci. Technol.* **14**, 722–733 (2005).

[62] Auriemma F., Corradini P. & Vacatello M. Order/disorder phase transitions of liquid-crystalline polymers with rigid groups in the side chains. A lattice theory. *J. Chem. Phys.* **93**, 8314–8320 (1990).

[63] Chen F. F. & Torreblanca H. Large-area helicon plasma source with permanent magnets. *Plasma Phys. Control. Fusion* **49**, A81–A88 (2007).

[64] Ashida S., Lee C. & Lieberman M. A. Spatially averaged (global) model of time modulated high density argon plasmas. *J. Vac. Sci. Technol. A* **13**, 2498–2507 (1995).

[65] Cho S. A self-consistent global model of neutral gas depletion in pulsed helicon plasmas. *Phys. Plasmas* **6**, 359–364 (1999).

[66] Wenxue Li, Yue Liu & Geng Wang. Damping characteristics of helicon and Trivelpiece–Gould waves in high density and low magnetic field helicon plasma. *Phys. Plasmas* **30**, 023510 (2023).

[67] Jiatong Wu, Chaofeng Sang, Chen Sun et al. Experimental and simulation study of argon helicon discharge in multiple plasma simulation linear device (MPS-LD). *Plasma Sources Sci. Technol.* **33**, 085007 (2024).